\documentclass[a4paper, 11pt]{article}
\usepackage{bm}
\usepackage{setspace}
\pdfoutput=1

\usepackage{makeidx}
\usepackage{amssymb}
\usepackage{amsmath}
\usepackage{amsfonts}
\usepackage{lineno}
\usepackage{epsfig}
\usepackage{graphics}
\usepackage{graphicx}
\usepackage{ntheorem}
\usepackage{color}
\usepackage[top=1in,bottom=1in,left=1.2in,right=1.2in]{geometry}
\usepackage{mathrsfs}
\usepackage{enumerate}
\usepackage{algorithm}
\usepackage{algorithmic}
\usepackage{float}
\usepackage{subcaption}
\usepackage{caption}
\usepackage{bbm}
\usepackage{algorithmic}

\usepackage{epstopdf}
\usepackage{subcaption}
\usepackage{tabularx}
\usepackage{booktabs}
\usepackage{textcomp}
\usepackage{xcolor}
\usepackage{mdframed} 
\usepackage{multirow}

\usepackage{stmaryrd}
\usepackage{epsfig}
\usepackage{xy}\xyoption{all}

\usepackage[bookmarks=false]{hyperref}
\usepackage{float}

\newtheorem{theorem}{Theorem}

\newtheorem{definition}{Definition}
\newtheorem{example}{Example}

\newtheorem{lemma}{Lemma}

\newtheorem{proposition}{Proposition}

\newcommand{\te}{\hfill $\Box$}

\newcommand{\be}{\begin{equation}}
\newcommand{\ee}{\end{equation}}
\newcommand{\ba}{\begin{eqnarray}}
\newcommand{\ea}{\end{eqnarray}}
\newcommand{\bas}{\begin{eqnarray*}}
\newcommand{\eas}{\end{eqnarray*}}
\def\text{\hbox}

\DeclareMathAlphabet{\mathbfit}{OML}{cmr}{bx}{it}

\begin{document}

\title{Private Markovian Equilibrium in Stackelberg Markov Games for Smart Grid Demand Response}

\author{Siying~Huang\thanks{S.Y.~Huang is with the School of Mathematical Sciences, University of Chinese Academy of Sciences, Beijing 100049, China (e-mail: huangsiying@amss.ac.cn). Y.F.~Mu and G.~Chen are with the State Key Laboratory of Mathematical Sciences(SKLMS), Academy of Mathematics and Systems Science, Chinese Academy of Sciences, Beijing 100190, China (e-mail: mu@amss.ac.cn; chenge@amss.ac.cn).}, Yifen~Mu\thanks{Corresponding author: Yifen Mu.}, Ge~Chen
}

%
%
\date{}
\maketitle

\begin{abstract}
The increasing integration of renewable energy introduces great challenge to 
the supply and demand balance of the power grid.  To address this challenge, this paper formulates a Stackelberg Markov game (SMG) between an aggregator and multiple users, where the aggregator sets electricity prices and users make demand and storage decisions. Considering that users’ storage levels are private information, we introduce private states and propose the new concepts of private Markovian strategies (PMS) and private Markovian equilibrium (PME).  
We establish the existence of a pure PME in the lower-level Markov game and prove that it can be computed in polynomial time. Notably, computing equilibrium in general Markov games is hard, and polynomial-time algorithms are rarely available.
Based on these theoretical results, we develop a scalable solution framework combining centralized and decentralized algorithms for the lower-level PME computation with upper-level pricing optimization. Numerical simulations with up to $50$ users based on real data validate the effectiveness and scalability of the proposed methods, whereas prior studies typically consider no more than $5$ users.
\begin{center}
\end{center}

\end{abstract}

%
%
\begin{spacing}{1.0}
\end{spacing}

\section{Introduction}
With the rapid development of renewable energy, particularly wind and solar power, large-scale integration of these sources into the smart grid has become an inevitable trend \cite{hossain2016role,bose2017power,sterl2020smart}. However, their inherent stochasticity, volatility, and intermittency significantly complicate both power dispatch and user behavior, posing serious challenges to maintaining the supply–demand balance and ensuring the secure and efficient operation of the grid \cite{mohandes2019review,smith2022effect,hossain2014renewable}.

Among various approaches to addressing these challenges, demand response (DR), especially real-time pricing (RTP), has been widely studied and recognized as an effective solution \cite{siano2014demand,li2021coordinating}. By dynamically adjusting electricity prices, RTP incentivizes users to shift consumption patterns, thereby improving load balancing and system efficiency \cite{xu2021hybrid}.
Moreover, energy storage technologies can effectively alleviate supply–demand fluctuations by allowing users to store energy when prices are low and use it when prices are high. Consequently, storage-enabled users have been considered in recent studies \cite{rahbar2014real, soliman2014game}.

In an RTP mechanism, there are typically two types of participants: aggregators, often operated by the state grid or third parties, and users such as factories, households, and commercial buildings. Their behaviors are tightly coupled, exhibiting clear game-theoretic interactions. Due to the hierarchical structure of the electricity market, where aggregator sets RTP and users respond by adjusting their consumption, this interaction naturally forms a \textit{Stackelberg game}, with the aggregator as the leader and users as followers \cite{saad2012game,zhao2017real,lu2018data,ding2020tracking}. 

In recent years, substantial progress has been made in Stackelberg game models. For example, \cite{maharjan2013dependable} considers multiple utility companies and users, where each user optimizes demand based on their own budget.
\cite{soliman2014game} analyzes storage-enabled users who respond to utility pricing while minimizing the same global cost function.
\cite{yu2015real} focuses on pricing and demand strategies between an energy management center and devices, deriving analytical solutions under DSC conditions.
\cite{jiang2022multi} examines the interaction between a distribution system operator and load aggregators across multiple timescales under rapid Photovoltaic generation fluctuations.

However, most of these works treat renewable generation as a deterministic process or do not include it at all. Thus, they are not suitable for modeling price-based demand response in the smart grid with large-scale renewable integration.

Recently, researchers have started to adopt Markov game (MG) model to better capture the uncertainty of renewable energy \cite{etesami2018stochastic,wu2024relationships}. \cite{etesami2018stochastic} mainly focuses on prosumers who make consumption and production decisions within an MG framework.  
\cite{wu2024relationships} also studies prosumers, but with more complex strategies and long-term discounted payoffs.
However, since computing Markovian equilibrium (ME) in MGs is generally difficult \cite{daskalakis2023complexity,deng2023complexity}, special structures are often assumed in the literature such as zero-sum games \cite{daskalakis2020independent,ozdaglar2021independent,sayin2022fictitious}, Markov potential games (MPGs) \cite{mguni2021learning,song2022when,fox2022independent,zhang2024gradient,leonardos2022global}, or independent chain games \cite{etesami2024learning}. Specifically, \cite{etesami2018stochastic} and \cite{wu2024relationships} assume that each player independently controls a Markov chain that characterizes their stochastic renewable generation.
In addition, some works consider relaxed solution concepts such as correlated and coarse correlated equilibrium (CE/CCE) \cite{zhang2022policy, erez2023regret}. To the best of our knowledge, these equilibrium concepts have not yet been applied to smart grid scenarios and are not the focus of this paper.
 
Motivated by the above discussion, in this paper, we propose a novel \textit{Stackelberg Markov Game} (SMG) framework to model the hierarchical and dynamic interactions between an aggregator and multiple users in a renewable-integrated smart grid with energy storage. To capture renewable uncertainty, we model renewable generation as a Markov chain, which influences both system pricing and user decision processes. Based on this, the upper-level aggregator (intermediary) sets prices according to system conditions, while lower-level users, considering their private and evolving storage levels, engage in a finite-horizon, non-cooperative Markov game. This framework breaks through the static or single-stage assumptions of traditional Stackelberg models and provides a more realistic depiction of decision-making under high dynamics and uncertainty in the smart grid.

In summary, this paper makes the main contributions as follows:

\begin{itemize}
\item We propose a SMG model that accounts for renewable energy uncertainty and user-side storage, providing a general framework for large-scale renewable integration. To our knowledge, this is the first work in demand response for smart grids to capture a hierarchical Stackelberg structure where the lower-level followers interact through a Markov game.
\item Considering that users’ storage levels are private and unobservable, we introduce private states and propose new concepts of private Markovian strategies (PMS) and the corresponding private Markovian equilibrium (PME). PMSs relax the full-observability assumption, providing a more realistic strategy for smart grids.
\item We establish the existence of a pure PME in the lower-level Markov game, and further strictly prove that  the pure PME can be computed in polynomial time. Notably, computing equilibrium in general Markov games is hard, and polynomial-time algorithms are rarely available.
\item We develop a solution framework for the SMG, combining lower-level PME computation (via a centralized MPG-based algorithm for complete information and a decentralized learning algorithm for incomplete information) with upper-level pricing optimization. Large-scale simulations with $50$ users demonstrate the effectiveness and scalability of the proposed methods, whereas prior studies typically consider no more than $5$ users \cite{etesami2018stochastic, hu2024economical}.
\end{itemize}

This paper is organized as follows.
Section~\ref{modelsection} presents the SMG model and introduces the concepts of PMS and the corresponding PME.
Section~\ref{proofsection} provides a constructive proof of PME existence in the lower-level MG and establishes an MPG whose equilibrium can be directly used to constructed the PME.
Section~\ref{algsection} develops a scalable equilibrium computation framework covering both lower-level PME computation and upper-level pricing optimization, with simulation validation on systems with up to 50 users. 
Finally, Section~\ref{conclusection} concludes the paper and outlines future directions.

\section{Model: Stackelberg Markov Game in renewable-integrated smart grid}\label{modelsection}

\begin{figure}[htb]
    \centering
    \includegraphics[width=\textwidth]{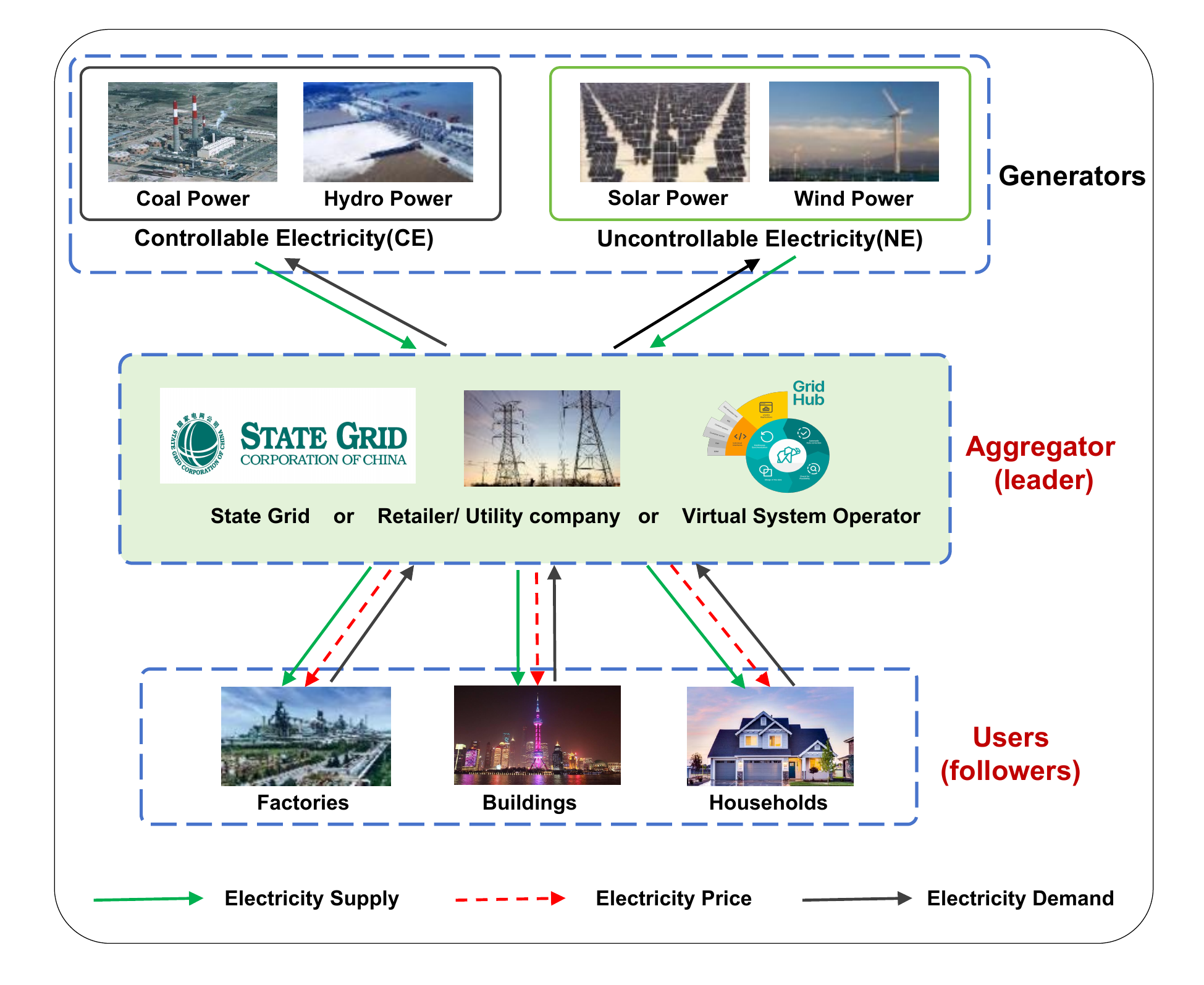}
    \caption{The conceptual framework of the electricity trading process.}
    \label{framework}
\end{figure}

In the renewable-integrated smart grid, there are three types of participants: the supply side, i.e., generators, producing electricity from coal, hydro power, wind, or solar sources; the demand side, i.e., the electricity users, such as factories, households, or commercial buildings; and the intermediary, also called the aggregator, which may be the state grid or third-party entities,  responsible for electricity transmission and distribution. See Fig.~\ref{framework} for an overview.
In this paper, we assume that the aggregator aims to minimize generation costs on the supply side, maintain supply–demand balance, prioritize renewable integration, and ensure stable system operation. Accordingly, we focus on the interactions between an aggregator and multiple storage users. To this end, the aggregator collects user demand, purchases electricity (prioritizing renewables), and sets the real-time price (RTP). Meanwhile, storage users optimize their decisions of consumption, demand, and storage based on the RTP. Fig.~\ref{framework} illustrates the conceptual framework of the electricity trading process.

Clearly, the interaction between an aggregator and multiple storage users is hierarchical. Therefore, we model the interaction as a multi-stage Stackelberg game, where the aggregator acts as the leader and multiple storage users as the followers.
The game is formalized below, and key notations are summarized in Table~\ref{notations}.
\begin{table}[ht]
\renewcommand{\arraystretch}{1.3}
\begin{center}\footnotesize
\setlength{\tabcolsep}{8pt}
\caption{LIST OF MAIN NOTATIONS.}
\label{notations}
\begin{tabularx}{\columnwidth}{c >{\raggedright\arraybackslash}X}
\toprule
\textbf{Notation} & \quad\quad\textbf{Description} \\
\midrule
$\mathcal{N}$ & Set of all users, $\mathcal{N}:=\{1,2,\dots,n\}$\\
$\mathcal{T}$ & Set of discrete time slots, $\mathcal{T}:=\{1,2,\dots,T\}$\\
$P(\cdot)$ & Unit electricity price function \\
$U(\cdot)$ & Payoff function of the leader\\
$e^t_{N}$ & Uncontrollable electricity at time $t$, $e^t_{N} = \hat{e}^t_{N} + \omega^t_{N} \in \mathcal{E}^t_{N}$ \\
$b^t_i$ & Energy storage level for user $i$ at time $t$, $b^t_i \in \mathcal{B}_i$ \\
$\boldsymbol{s}^t$ & State at time $t$, $\boldsymbol{s}^t = (e^t_{N}, b^t_1, \dots, b^t_n) \in \mathcal{S}^t$ \\
$\boldsymbol{s}_i^t$ & Private state of user $i$ at time $t$, $\boldsymbol{s}_i^t = (e^t_{N}, b^t_i) \in \mathcal{S}_i^t$ \\
$d^t_i$ & Electricity demand of user $i$ at time $t$, $d^t_i \in \mathcal{D}_i$ \\ 
$c^t_i$ & Electricity consumption of user $i$,  $c^t_i \in \mathcal{C}_i$ \\
$a^t_i = (d^t_i, c^t_i)$ & Action of user $i$ at time $t$, $a^t_i \in \mathcal{A}^t_i = \mathcal{D}_i \times \mathcal{C}_i$ \\
$r_i(\boldsymbol{s}^t,\boldsymbol{a}^t)$ & Stage payoff of user $i$ \\
$q(\cdot \mid \boldsymbol{s}^t, \boldsymbol{a}^t)$ & Transition distribution of the next state \\
$\boldsymbol{\sigma}_i$ & Markovian strategy of user $i$, $\boldsymbol{\sigma}_i \in \Sigma_i$, with $\boldsymbol{\sigma}_i = (\sigma_i^1,\dots,\sigma_i^T)$, $\sigma_i^t: \mathcal{S}^t \to \Delta(\mathcal{A}_i^t)$ \\
$\boldsymbol \pi_i$ & PMS of user $i$, $\boldsymbol \pi_i \in \Pi_i$, with $\boldsymbol{\pi}_i = (\pi_i^1,\dots,\pi_i^T)$, $\pi_i^t: \mathcal{S}_i^t \to \Delta(\mathcal{A}_i^t)$ \\
$\boldsymbol \pi$ & PMS profile of all users,  $\boldsymbol \pi \in \Pi$ \\
$V^{\boldsymbol \pi}_{t,i}(\boldsymbol{s}^t)$ & Value function of user $i$ under strategy profile $\boldsymbol \pi$ \\
$\phi_{\boldsymbol{s}}(\cdot)$ & State-dependent potential function \\
\bottomrule
\end{tabularx}
\end{center}
\end{table}

\subsection{Leader (the Aggregator) Model}
As the leader, the aggregator purchases electricity to meet users' demand and determines the RTP mechanism to sell electricity to users.

First, the electricity bought by the aggregator from the supply side consists of two parts:
\begin{enumerate}
\item \label{item 1} \textit{Uncontrollable electricity} ($e^t_N$):  
    This refers to renewable generation (e.g., wind or solar), which is inherently stochastic due to weather variability. At time $t$, the predicted output is $\hat{e}^t_N$, and the forecasting error is $\omega^t_N := e^t_N - \hat{e}^t_N$. We assume $\omega^t_N$ follows a Markov chain with transition probability $\hat{q}(\omega^{t+1}_N \mid \omega^t_N)$:
    \begin{align}\label{hat q}
    \hat{q}(\omega^{t+1}_N \mid \omega^t_N) = \hat{q}(\omega^{t+1}_N \mid \omega^1_N, \dots, \omega^t_N).
    \end{align}
    
    Since $e^t_N$ $=$ $\hat{e}^t_N$ $+$ $\omega^t_N$ and $\omega^t_N$ is Markovian, the sequence $\{e^t_N\}_{t=1}^T$ also forms a Markov chain, with transition probability likewise denoted by $\hat{q}$. At the beginning of the game, the forecast sequence $\{\hat{e}_N^t\}_{t=1}^T$ and the Markov transition probability $\hat{q}$ are assumed to be known and fixed. This makes it suitable for modeling renewable uncertainty \cite{meyn2015ancillary}.
    \item \textit{Controllable electricity} ($e^t_C$): This includes dispatchable sources such as coal or hydro. To prioritize the use of renewable energy and maintain real-time supply-demand balance, controllable generation is used only to meet the residual demand. Following standard assumptions, we impose:
    \begin{align*}
         e_{C}^t = d^t - {e}_{N}^t,\quad \forall t\in \mathcal{T},
    \end{align*}
    where $ d^t = \sum_{i=1}^{n} d_{i}^t$ denotes the total electricity demand of users at time $t$. 
\end{enumerate}

Second, to sell electricity, the aggregator announces RTP rules to energy users. Following many studies that adopt linear RTP due to its simplicity and effectiveness in capturing key price–response dynamics \cite{liu2017optimal,wang2015optimal}, we use the following pricing function:
\begin{align}\label{price}
P( d^t, e_{N}^t)=\frac{\alpha}{n  {e}_{N}^t+\gamma_{1}} d^t+\frac{\beta} { e_{N}^t+\gamma_{2}}, 
\end{align}
where  $\alpha,\beta, \gamma_{1}, \gamma_{2}>0$ are adjustable parameters chosen by the aggregator. By \eqref{price}, price depends on actions $\alpha,\beta, \gamma_{1}$ and $\gamma_{2}$ of the aggregator/leader, the total demand $ d^t$ of users/followers, and uncontrollable electricity $e_{N}^t$. This dependency makes the RTP an effective method for demand response. 

The aggregator's total \textit{payoff} is defined as:
\begin{align}\label{aggregator_payoff}
U = \mathbb{E}_{\{e_{N}^t\}}\Big[  \sum_{t=1}^{T} \Big( 
     P(d^t, e_{N}^t)  d^t - C  (d^t - e_{N}^t)  - \frac{1}{2} k (d^t - e_{N}^t - r_0)^2 
\Big)\Big].
\end{align}
Here, $C>0$ denotes the unit cost of controllable generation, $k>0$ is the penalty weight, and $r_0$ is the target level of controllable electricity used to penalize deviations. This payoff consists of three components: i) revenue from electricity sales; ii) cost of controllable electricity generation; and iii) a penalty for instability in controllable electricity. Here, we assume that renewable generation has negligible cost to highlight its advantage. By maximizing \eqref{aggregator_payoff}, the aggregator aims to balance system stability and generator profits, thus maximizing social welfare.

We now turn to the strategic interactions among energy storage users.

\subsection{Followers (Energy Storage Users) Model: Markov Games and Preliminaries}
Given the RTP mechanism set by the aggregator, the strategic interactions between users form a finite-horizon non-cooperative Markov game (MG) \cite{shapley1953stochastic}, denoted as 
\begin{align}\label{G1}
\mathcal{G}_1 = \left\{\mathcal{T},\mathcal{N},\{\mathcal{S}^t\}_{t=1}^T,\left\{\mathcal{A}_{i}\right\}_{i=1}^{n},\left\{{r}_{i}(\cdot)\right\}_{i=1}^{n},q(\cdot \mid \cdot ) \right\}, 
\end{align}
where $\mathcal{T}, \mathcal{N}$, and $\{\mathcal{S}^t\}_{t=1}^T$ denote the time slot set, user set, and the collection of stage-wise state spaces, respectively;
$\mathcal{A}_{i}$ and $r_i(\cdot)$ denote the action set and payoff function of user $i$; and $q(\cdot \mid \cdot )$ is the state transition probability function. Their detailed definitions are listed as follows:

\begin{enumerate}
	\item The state space at time $t$ $\mathcal{S}^t := \mathcal{E}_N^t \times \prod_{i=1}^n \mathcal{B}_i$ consists of both public and private components: the first part $\mathcal{E}_N^t$ is the finite set of all possible values of uncontrollable electricity $e^t_{N}$ (defined in item \ref{item 1}), which is public information, and the second part  $\mathcal{B}_i:=\{ 0,1, \dots, b^{\text{max}}_{i}\}$ represents the set of all possible energy storage levels of user $i$, which is private information.
	\item The action set of user $i$ is defined as 
    $\mathcal{A}_{i}:=\prod_{t\in \mathcal{T}} \mathcal A^t_i := \prod_{t\in \mathcal{T}} \left\{a^t_{i} = (d^t_{i},c^t_{i}) \in \mathcal D_{i} \times \mathcal C_{i} \right\},$ where $a_i^t$ denotes the action of user $i$ at time $t$, consisting of electricity demand $d_i^t$ and consumption $c_i^t$, with $d_i^t \in \mathcal{D}_i := \{0, 1, \dots, d_i^{\text{max}}\}$ and $c_i^t \in \mathcal{C}_i := \{0, 1, \dots, c_i^{\text{max}}\}$. These two variables are constrained by the stored electricity level $b_i^t$:
    \begin{align}\label{relationship} b^t_{i} +  d^t_{i} - b^{\text{max}}_{i}\leq c^t_{i} \leq b^t_{i} +  d^t_{i}.\end{align} 
    To ensure the feasibility of constraint~\eqref{relationship}, we assume $c_i^{\text{max}}$ is sufficiently large. Denote $\mathcal{A}^t:=\prod_{i\in \mathcal{N}}\mathcal A^t_i$ as the action profile set of all users at time $t$. 
	\item $r_i(\boldsymbol{s}^t,\boldsymbol{a}^t)$ is the stage payoff of user $i$, which depends on the current system state $\boldsymbol{s}^t=(e^t_{N},b^t_1,\dots,b^t_n)\in \mathcal{S}^t$ and the joint action $\boldsymbol{a}^t=(a^t_{1},a^t_{2},\dots,a^t_{n})\in \mathcal{A}^t$ of all users. Throughout this paper we set 
 \begin{align}\label{current reward}
r_i(\boldsymbol{s}^t,\boldsymbol{a}^t) :=  \theta_{i} c_{i}^{t} -P(d^{t}, e^t_{N})  d^{t}_{i},
 \end{align}
where $\theta_i > 0$ is the consumption benefit coefficient of user~$i$.
The payoff consists of the benefit from electricity consumption minus the cost of purchasing electricity.
The linear form of the consumption benefit is a commonly adopted and reasonable simplification \cite{eksin2016demand}, \cite{muthirayan2019mechanism}.  Note that, given the state $\boldsymbol{s}^t$ and the pricing function \eqref{price}, the stage payoff $r_i(\boldsymbol{s}^t, \boldsymbol{a}^t)$ exhibits an aggregative structure: it depends only on user $i$’s own action and the aggregate demand $d^t_{-i} = \sum_{j \ne i} d^t_j$ of other users.
 
	\item The state transition probability $q(\cdot \mid \boldsymbol{s}^t, \boldsymbol{a}^t)$ consists of two components. First, the uncontrollable electricity $e_N^t$ evolves as the Markov chain with transition probability $\hat{q}(\cdot)$ \eqref{hat q}, independent of users’ actions. Second, the storage level of each user updates deterministically given their current state and actions, according to: $b_i^{t+1} = b_i^t + d_i^t - c_i^t, \forall i \in \mathcal{N}$. These two components jointly yield the following transition:
\begin{align}\label{q}
&q(\boldsymbol s^{t+1} = (\hat e^{t+1}_{N} + \omega^{t+1}_{N}, b^{t+1}_1, \dots,b^{t+1}_n) \mid  \boldsymbol s^t = (\hat e^t_{N} + \omega^t_{N}, b^t_1,\dots,b^t_n),\boldsymbol a^t) \nonumber\\
& = 
\hat q(\omega^{t+1}_{N} \mid \omega^{t}_{N}).
\end{align}
Here we recall that $\hat e^{t}_{N}$ and $\omega^{t+1}_{N}$ are the predicted value and error of $e^{t}_{N}$ respectively, which means $e^{t}_{N} = \hat e^{t}_{N} + \omega^{t}_{N}$.
\end{enumerate}

In finite-horizon MGs, it is standard to assume that users adopt Markovian strategies \cite{maschler2020game}:
\begin{definition}[Markovian strategy]\label{GMS} For any player $i$, if its strategy $\boldsymbol{\sigma}_i(\cdot):=(\sigma_i^1(\cdot),\ldots,\sigma_i^T(\cdot))$ satisfies 
$\sigma_i^t(\boldsymbol{s}^t)\in\Delta(\mathcal{A}_{i}^t)$ for all $t\in\mathcal{T}, \boldsymbol{s}^t\in\mathcal{S}^t$, 
where $\sigma_i^t(\cdot)$ is the stage-$t$ decision rule of player $i$, mapping the state 
$\boldsymbol{s}^t$ to a distribution over $\mathcal{A}_i^t$, 
we say $\boldsymbol{\sigma}_i$ is a Markovian strategy of player $i$.
\end{definition}

Here, $\Delta(\mathcal{A}_i^t)$ denotes the probability simplex over $\mathcal{A}_i^t$. Let $\Sigma_i$ denote the set of all such Markovian strategies of player $i$ and we will write $\boldsymbol \sigma = (\boldsymbol \sigma_i)_{i\in \mathcal{N}} \in \Sigma := \times_{i\in \mathcal{N}} \Sigma_{i}, \boldsymbol \sigma_{-i} = (\boldsymbol \sigma_j)_{i \neq j \in \mathcal{N}} \in \Sigma_{-i} := \times_{i \neq j \in \mathcal{N}} \Sigma_{j}$ to denote the Markovian strategies profile of all players and of all players other than $i$, respectively. In other words, a Markovian strategy determines actions based solely on the current time $t$ and state $\boldsymbol s^t$, without considering the full history. 

The value function $V^{\boldsymbol \sigma}_{t,i}(\boldsymbol s): \mathcal{S}^t \rightarrow \mathbb{R}$ is defined as expected cumulative reward for player $i$ when Markovian strategy profile $\boldsymbol \sigma$ is taken starting from state $\boldsymbol s$ and time $t$:
\begin{align}\label{V-function}
    V^{\boldsymbol \sigma}_{t,i}(\boldsymbol s) := \mathbb{E} \Big[\sum_{t^{\prime} = t}^{T} r_{i}\left(\boldsymbol s^{t^{\prime}}, \boldsymbol a^{t^{\prime}}\right) \mid \boldsymbol s^t = \boldsymbol s\Big],
\end{align}
where the expectation is with respect to the random trajectory $\tau = (t^{\prime},\boldsymbol s^{t^{\prime}}, \boldsymbol a^{t^{\prime}}, r_i)_{t^{\prime}=t}^{T}$, where ${a}^{t^{\prime}}_i \sim  \sigma_i^{t^{\prime}}(\boldsymbol{s}^{t^{\prime}})$, $\boldsymbol{s}^{t^{\prime}+1} \sim P(\cdot \mid \boldsymbol{s}^{t^{\prime}}, \boldsymbol{a}^{t^{\prime}})$.

Markovian equilibrium (ME) \cite{maschler2020game} is the conventional solution concept under Markovian strategies. In the single-stage case ($T = 1$), the ME reduces to a Nash equilibrium of the resulting strategic-form game.
\begin{definition}[Markovian equilibrium (ME)]\label{ME}
A Markovian strategy profile $\boldsymbol \sigma^{*}$ is an ME if 
\begin{align*}
    V_{1,i}^{\boldsymbol \sigma^{*}_{i}, \boldsymbol \sigma^{*}_{-i}}(\boldsymbol s^1) \geq V_{1,i}^{\boldsymbol \sigma_{i}, \boldsymbol \sigma^{*}_{-i}}(\boldsymbol s^1) , ~ \forall \boldsymbol s^1 \in \mathcal{S}^1, i \in \mathcal{N}, \boldsymbol \sigma_{i} \in \Sigma_{i}.
\end{align*}
\end{definition}

\subsection{Private Markovian Strategy (PMS) and Private Markovian Equilibrium (PME)}
In Definition~\ref{GMS}, Markovian strategies are defined under the assumption of full state observability. However, in the smart grid model, each user's storage level is private and unobservable to others. To capture this private information structure, we introduce a more realistic strategy: the private Markovian strategy (PMS), which is a special Markovian strategy.

At each time $t$, user~$i$ observes a private state $\boldsymbol{s}_i^t := (e_N^t, b_i^t) \in \mathcal{S}_i^t := \mathcal{E}_N^t \times \mathcal{B}_i$, consisting of the uncontrollable electricity $e_N^t$ (as public information) and the individual storage level $b_i^t$ (as private information). A PMS restricts user~$i$’s actions to depend only on the current time and private state:
\begin{definition}[Private Markovian strategy (PMS)]\label{PMS}
For any player $i$, if its strategy  $\boldsymbol{\pi}_i(\cdot):=(\pi_i^1(\cdot),\ldots,\pi_i^T(\cdot))$ satisfies $\pi_i^t(\boldsymbol{s}_i^t)\in\Delta (\mathcal{A}_{i}^t)$ for all $t \in \mathcal{T}, \boldsymbol{s}_i^t \in \mathcal{S}^t_i$, where $\pi_i^t(\cdot)$ is the stage-$t$ decision rule of player $i$, mapping its private state $\boldsymbol{s}_i^t$ to a distribution over $\mathcal{A}_i^t$, we say $\boldsymbol{\pi}_i$ is a PMS of player $i$.
Specifically, a pure PMS of player $i$ is a deterministic mapping $\boldsymbol{\mathbbm{1}}_{\boldsymbol{a}_i}(\cdot):=(\mathbbm{1}_{a_i^1}(\cdot),\ldots,\mathbbm{1}_{a_i^T}(\cdot))$ satisfying $\mathbbm{1}_{a_i^t}(\boldsymbol{s}_i^t) \in \mathcal{A}_i^t$ for all $t \in \mathcal{T}, \boldsymbol{s}_i^t \in \mathcal{S}^t_i$, where $\boldsymbol{a}_i=(a_i^1,\ldots,a_i^T).$
\end{definition}

Let $\Pi_i$ denote the set of all PMSs of player $i$, and $\mathcal{P}^{PMS}_i$ the set of all pure PMSs of player $i$. We will write $\boldsymbol \pi = (\boldsymbol \pi_i)_{i\in \mathcal{N}} \in \Pi := \times_{i\in \mathcal{N}} \Pi_{i}$, $\boldsymbol \pi_{-i} = (\boldsymbol \pi_j)_{i \neq j \in \mathcal{N}} \in \Pi_{-i} := \times_{i \neq j \in \mathcal{N}} \Pi_{j}$ to denote the PMS profile of all players and of all players other than $i$, respectively.

Any PMS corresponds to a Markovian strategy that assigns the same action distribution to all full states sharing the same private state, but the converse may not hold.

\begin{example}[PMS as a Special Markovian Strategy]
Consider a two-player MG where the full state at time $t$ is $\boldsymbol{s}^t = (e^t_N, b^t_1, b^t_2)$, with $e^t_N = 5$, $b^t_1 = 1$, and $b^t_2 \in \{1,2\}$.
For player~1, both states $(5,1,1)$ and $(5,1,2)$ correspond to the same private state $(5,1)$.
A PMS for player~1 must assign the same action distribution $\pi_1^t(5,1)$ to both states,
whereas a general Markovian strategy $\boldsymbol \sigma_1$ may assign different distributions $\sigma_1^t(5,1,1)$ and $\sigma_1^t(5,1,2)$.
\end{example}

Given a PMS profile $\boldsymbol \pi$, the value function of player $i$ starting from state $\boldsymbol s \in \mathcal S^t$ and time $t$ is defined as:
\begin{align}\label{V_PMS}
    V^{\boldsymbol \pi}_{t,i}(\boldsymbol s) := \mathbb{E} \Big[\sum_{t^{\prime} = t}^{T} r_{i}\left(\boldsymbol s^{t^{\prime}}, \boldsymbol a^{t^{\prime}}\right) \mid \boldsymbol s^t = \boldsymbol s\Big].
\end{align}
Note that although the definition of $V^{\boldsymbol \pi}_{t,i}(\boldsymbol s)$ shares the same form as in \eqref{V-function}, the expectation here is taken over actions sampled from the PMS, i.e., $a^{t^{\prime}}_i \sim \pi_i^{t^{\prime}}(\boldsymbol{s}_i^t)$ and the state transitions $\boldsymbol s^{t^{\prime}+1}\sim P(\cdot \mid \boldsymbol s^{t^{\prime}}, \boldsymbol a^{t^{\prime}})$.

Then, we define the \textit{private Markovian equilibrium}:
\begin{definition}[Private Markovian equilibrium (PME)]\label{PME}
A PMS profile $\boldsymbol \pi^{*}$ is a PME if 
\begin{align*}
    V_{1,i}^{\boldsymbol \pi^{*}_{i},\boldsymbol \pi^{*}_{-i}}(\boldsymbol s^1) \geq V_{1,i}^{\boldsymbol \pi_{i}, \boldsymbol \pi^{*}_{-i}}(\boldsymbol s^1), ~\forall \boldsymbol{s}^1 \in \mathcal{S}^1,i \in \mathcal{N},\boldsymbol \pi_{i} \in \Pi_{i}.
\end{align*}
\end{definition}

\vskip 2mm

PME generalizes ME to partially observable settings, where strategies rely on both public and private information. This extension is particularly relevant in practical systems such as smart grids (where user storage levels are private) or epidemic modeling (where individual health status are private). 

\begin{figure}[htb]
    \centering
    \includegraphics[width=0.9\textwidth]{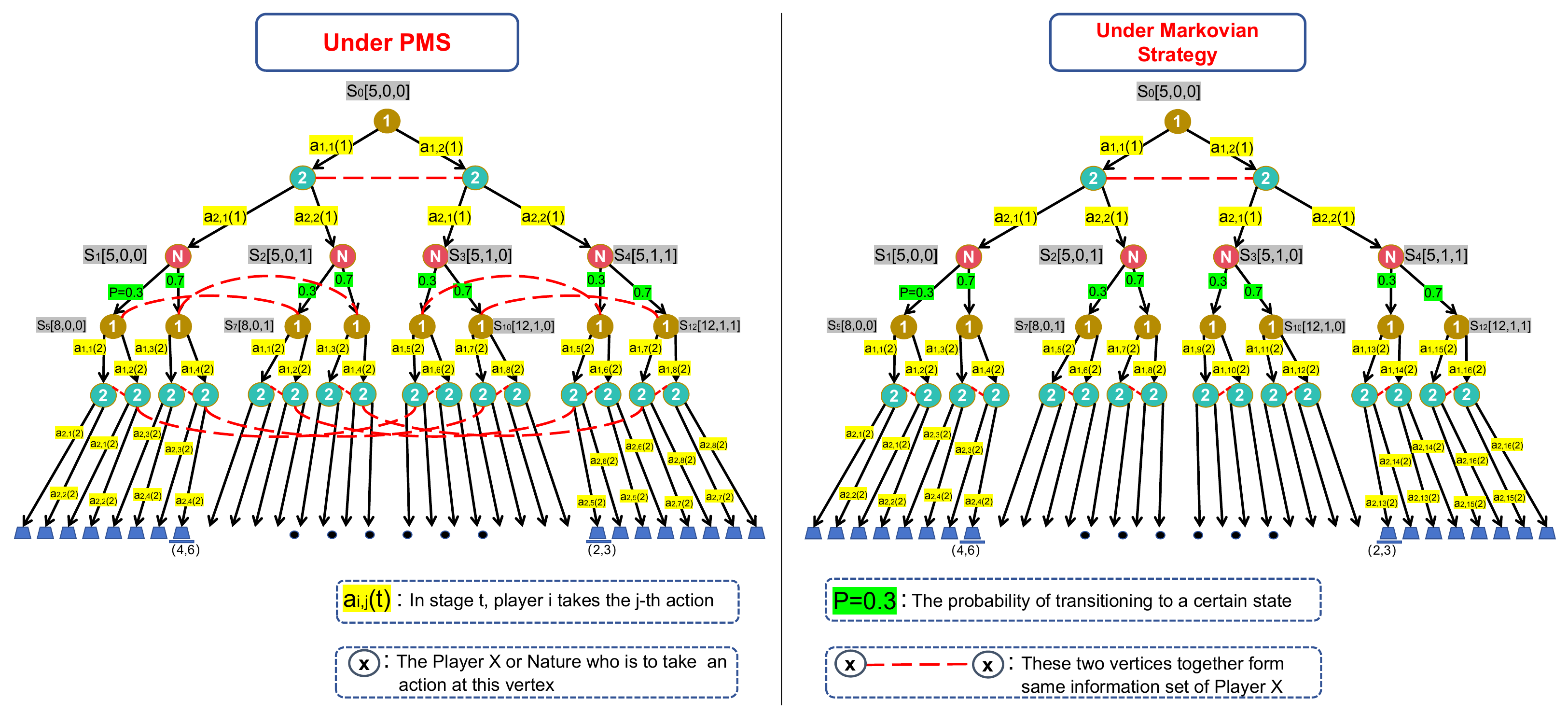} 
    \caption{Game tree of a two-player, two-horizon Markov game starting from $\boldsymbol{s}_0 = [5, 0, 0]$, where the initial uncontrollable electricity generation is 5 and both players have zero storage.
At stage~1, players select actions based on their private state under PMS, or on the full state under Markovian strategies.
The system transitions deterministically to states $s_1,\dots,s_4$ (nodes labeled ‘N’), followed by a stochastic update of renewable generation, leading to $s_5,\dots,s_{13}$.
At stage~2, players act again under PMS or Markovian strategies.
Under PMS, private states with identical local observations (storage and renewable level) form an information set, leading to consistent actions; such nodes are linked by red dashed lines.
In contrast, under Markovian strategies, all states are fully distinguishable and each node forms a singleton information set. 
The game ends with total payoffs $(\text{payoff}_1, \text{payoff}_2)$ at terminal nodes.}
    \label{game tree}
\end{figure}

To illustrate how the Markov game unfolds under both PMS and Markovian strategies, Fig.~\ref{game tree} depicts a two-player, two-horizon Markov game. It highlights sequential decisions and information sets, revealing structural differences between PMS and Markovian strategies.

We have now formulated the SMG, with the lower-level interactions modeled as an MG $\mathcal{G}_1$ with private information.

To solve the leader’s optimal pricing problem, we need to compute the equilibrium of the lower-level MG $\mathcal{G}_1$, as the leader’s strategy depends on users’ equilibrium responses. However, finding an ME in general-sum Markov games is PPAD-complete~\cite{daskalakis2023complexity,deng2023complexity}, making the problem computationally challenging.

In our setting, MG $\mathcal{G}_1$ involves private information on the users’ side. We therefore focus on PMS, which is a special case of the Markovian strategies introduced earlier. Nonetheless, the existence and computation of a PME remain nontrivial in general MGs. The existence of PME and its connection to the standard ME are discussed in Appendix~\ref{appendix_pme} of this paper.

These challenges motivate a deeper investigation into the structural properties of $\mathcal{G}_1$, which is the focus of the next section.

\section{Theoretical Results}\label{proofsection}
We begin by showing that the specific structure of MG $\mathcal{G}_1$ admits a constructive proof for the existence of a pure PME. Then, we demonstrate that the PME can be computed in polynomial time by converting the PME problem into the ME problem of a constructed MPG, thereby revealing an underlying potential structure that enables tractable equilibrium computation.

To this end, we first present the definitions of potential games and their Markov extensions (MPGs), which will be used in the conversion analysis that follows.

\subsection{Definitions of Potential and Markov Potential Games}

First, we introduce (exact) potential games, as defined by Shapley in 1996 \cite{monderer1996potential}.
\begin{definition}[Potential game]\label{Potential Game} 
A strategic-form game $\langle \mathcal{N}$, $(\mathcal{X}_i)_{i \in \mathcal{N}}$, $ (u_i)_{i \in \mathcal{N}}\rangle$ is a \textit{potential game
of pure strategies} if there exists a potential function $\phi: \mathcal{X} \rightarrow \mathbb{R}$ such that
\begin{align}\label{potential game function}
    u_{i}\left(x_{i}, x_{-i}\right)-u_{i}\left(x_{i}^{\prime}, x_{-i}\right)=\phi\left(x_{i}, x_{-i}\right)-\phi\left(x_{i}^{\prime}, x_{-i}\right), \notag \\
    \forall  i \in \mathcal{N}, x_{i}, x_{i}^{\prime} \in \mathcal{X}_{i}, x_{-i} \in \mathcal{X}_{-i},
\end{align}
and is a \textit{potential game
of mixed strategies} if there exists a potential function $\phi: \mathcal{M} \rightarrow \mathbb{R}$ such that 
\begin{align}\label{potential game function 1}
    u_{i}\left(m_{i}, m_{-i}\right)-u_{i}\left(m_{i}^{\prime}, m_{-i}\right)=
     \phi\left(m_{i}, m_{-i}\right)-\phi\left(m_{i}^{\prime}, m_{-i}\right), \notag \\
      \forall i \in \mathcal{N}, m_{i}, m_{i}^{\prime} \in \mathcal{M}_{i}, m_{-i} \in \mathcal{M}_{-i}.
\end{align}
\end{definition}

In the above definition, $m_i \in \mathcal{M}_{i}$ represents the mixed strategy of player $i$, while $m \in \mathcal{M}$ and $m_{-i} \in \mathcal{M}_{-i}$ denote the mixed strategies profile of all players and of all players other than $i$, respectively. The pure and mixed strategy formulations are equivalent, as they induce the same potential-maximizing behavior.

The concept of potential games can be naturally extended to the Markov game setting, leading to the notion of MPGs \cite{song2022when,fox2022independent}.
\begin{definition}[Markov potential game (MPG)]\label{Stochastic Potential Game 1}
A finite-horizon MG 
$\{\mathcal{T},$ $\mathcal{N},$ $ \{\mathcal{S}^t\}_{t=1}^T,$ $\left\{\mathcal{A}_{i}\right\}_{i=1}^{n},$ $\left\{{r}_{i}\right\}_{i=1}^{n}, q\}$  is an \textit{MPG
of pure Markovian strategies} if there exists a state-dependent potential function $\phi_{\boldsymbol{s}}: \mathcal{P}^{MS} \rightarrow \mathbb{R}$ such that 
\begin{align}\label{Stochastic potential game function 1}
V^{\boldsymbol{\mathbbm{1}}_{\boldsymbol{a}_{i}},\boldsymbol{\mathbbm{1}}_{\boldsymbol{a}_{-i}}}_{1,i}(\boldsymbol s^{1}) -V^{\boldsymbol{\mathbbm{1}}_{\boldsymbol{a}_i^{\prime}},\boldsymbol{\mathbbm{1}}_{\boldsymbol{a}_{-i}}}_{1,i}(\boldsymbol s^{1})= \phi_{\boldsymbol s^{1}}\left(\boldsymbol{\mathbbm{1}}_{\boldsymbol{a}_{i}},\boldsymbol{\mathbbm{1}}_{\boldsymbol{a}_{-i}}\right) -\phi_{\boldsymbol s^{1}}\left(\boldsymbol{\mathbbm{1}}_{\boldsymbol{a}_i^{\prime}},\boldsymbol{\mathbbm{1}}_{\boldsymbol{a}_{-i}}\right),\notag \\
\forall \boldsymbol s^{1} \in \mathcal S^1, i \in \mathcal{N}, \boldsymbol{\mathbbm{1}}_{\boldsymbol{a}_{i}}, \boldsymbol{\mathbbm{1}}_{\boldsymbol{a}_i^{\prime}} \in \mathcal{P}_i^{MS}, \boldsymbol{\mathbbm{1}}_{\boldsymbol{a}_{-i}} \in \mathcal{P}_{-i}^{MS},
\end{align}
and is an \textit{MPG
of general Markovian strategies} if there exists a state-dependent potential function $\phi_{\boldsymbol{s}}: \Sigma \rightarrow \mathbb{R}$ such that 
for every $\boldsymbol s^{1} \in \mathcal S^1, i \in \mathcal{N},\boldsymbol \sigma_i, \boldsymbol \sigma_i^{\prime} \in \Sigma_{i}$ , and  $ \boldsymbol \sigma_{-i} \in \Sigma_{-i}$,
\begin{align}\label{Stochastic potential game function}
    V^{\boldsymbol \sigma_i,\boldsymbol \sigma_{-i}}_{1,i}(\boldsymbol s^{1}) -V^{\boldsymbol \sigma_i^{\prime},\boldsymbol \sigma_{-i}}_{1,i}(\boldsymbol s^{1})= \phi_{\boldsymbol s^{1}}\left(\boldsymbol \sigma_i,\boldsymbol \sigma_{-i}\right)-\phi_{\boldsymbol s^{1}} \left(\boldsymbol \sigma_i^{\prime},\boldsymbol \sigma_{-i}\right),\notag \\
    \forall \boldsymbol s^{1} \in \mathcal S^1, i \in \mathcal{N}, \boldsymbol \sigma_i, \boldsymbol \sigma_i^{\prime} \in \Sigma_{i}, \boldsymbol \sigma_{-i} \in \Sigma_{-i}.
\end{align}
\end{definition}
Here $\boldsymbol{\mathbbm{1}}_{\boldsymbol{a}_i} \in \mathcal{P}_i^{MS}$ denotes a pure Markovian strategy of player $i$, 
written in the same stage-wise form as pure PMS in \ref{PMS} but defined on states $\boldsymbol{s}^t \in \mathcal{S}^t$. 
The notation is consistent with pure PMS, while the admissible set is $\mathcal{P}_i^{MS}$. The corresponding value function $V^{\boldsymbol{\mathbbm{1}}_{\boldsymbol{a}_{i}},\boldsymbol{\mathbbm{1}}_{\boldsymbol{a}_{-i}}}_{t,i}(\boldsymbol s)$ is computed via \eqref{V-function}.

The general Markovian strategies formulation~\eqref{Stochastic potential game function} in Definition~\ref{Stochastic Potential Game 1} aligns with standard definitions in the literature~\cite{song2022when}. In this work, we propose a novel pure strategy formulation~\eqref{Stochastic potential game function 1} for finite-horizon MG settings and prove its equivalence to the general Markovian strategies formulation (see Lemma \ref{equivalent of potential definition} in Appendix \ref{appendix2}). This alternative enables more tractable verification in applications.

A key property of finite-horizon MPGs is the guaranteed existence of a pure ME \cite{song2022when}, \cite{monderer1996potential}, \cite{chew2016potential}. This equilibrium can be obtained by maximizing the potential function in \eqref{Stochastic potential game function 1}, i.e., 
$(\boldsymbol{\mathbbm{1}}_{\boldsymbol{a}^*_i},\boldsymbol{\mathbbm{1}}_{\boldsymbol{a}^*_{-i}}) = \arg\max_{\boldsymbol{\mathbbm{1}}_{\boldsymbol{a}_i},\boldsymbol{\mathbbm{1}}_{\boldsymbol{a}_{-i}}} \phi_{\boldsymbol{s}^1}(\boldsymbol{\mathbbm{1}}_{\boldsymbol{a}_i},\boldsymbol{\mathbbm{1}}_{\boldsymbol{a}_{-i}}), \forall \boldsymbol s^{1} \in \mathcal S^1$.

For completeness, an extension of the definition of MPGs under the PMS framework is provided in Appendix~\ref{appendixX}.

\subsection{Existence and Computation of Pure PME}
With the above definitions in place, we begin with the single-stage case, where user decisions depend only on the current state $\boldsymbol s^t$. In this setting, the game simplifies to a strategic-form game without temporal coupling, and exhibits the following structural property and simplified equilibrium characterization:

\begin{lemma}\label{one-horizon}
Consider the single-stage MG ($T=1$) with a given state $\boldsymbol s^{t}= (e^t_{N}, b^t_1, \dots, b^t_n) \in \mathcal S^t$. Then:
\begin{enumerate}
\item \label{prop:exact-potential} the MG is an exact potential game;
\item \label{prop:equilibrium-only-en} the pure Nash equilibrium demand \( d^{*t}_i \) for each user \( i \in \mathcal{N} \) depends only on \( e^t_N \); that is, $d^{*t}_i(\boldsymbol s^{t})= d^{*t}_i(e^t_N)$.
\end{enumerate}
\end{lemma}
The proof of Lemma \ref{one-horizon} is postponed to Appendix \ref{prop1}.

Building on the single-stage analysis, we extend the analysis to the full $T$-horizon MG $\mathcal{G}_1$. Without loss of generality, we assume that all users start with zero initial energy storage levels, i.e., $\boldsymbol{s}^1 = (e^1_{N}, b^1_1=0, \dots, b^1_n=0) \in \mathcal{S}^1$. This assumption reflects many real-world scenarios (e.g., new users joining the system) and does not affect the generality of our analysis.
\begin{theorem}[Existence of pure PME in Markov game $\mathcal{G}_1$]\label{main thm}
The $T$-horizon Markov game $\mathcal{G}_1$ has a pure PME.
\end{theorem}
\textit{Proof sketch:} We apply backward induction to construct the pure PME of the Markov game $\mathcal{G}_1$ in four steps.

\textit{Step 1.} At stage \(T\), strictly dominated strategies are first eliminated, and the stage-game equilibrium then is obtained by Lemma~\ref{one-horizon}. 

\textit{Step 2.} Based on the partial independence of state transitions and the structure of the stage-\(T\) payoff (one part depending only on the renewable state, the other only on the individual storage level), the expected payoff at stage \(T\) can be simplified. 

\textit{Step 3.} At stage \(T-1\), the stage payoff is combined with the simplified expected payoff from stage \(T\), resulting in a recursive equation for the value function.

\textit{Step 4.} By recursively repeating Steps 1–3 at each earlier stage, we apply backward induction to obtain the pure PME.

The detailed proofs are provided in Appendix \ref{thm_1}.

Although a pure PME can be obtained through the recursive backward induction method described above, the computation becomes intensive since it requires solving equilibria across a large number of subgames.

To reduce complexity, we exploit the structural insights from Theorem~\ref{main thm}: i) after eliminating strictly dominated actions, each user’s decision reduces to choosing a demand $d^t_i$ depending only on $e^t_N$; and ii) the recursive value function admits a separable form as in \eqref{eq:value_function}. Based on these observations, we establish the following result.

\begin{theorem}[Polynomial-time PME Computation]\label{thm:MPG_equivalence}
For the $T$-horizon MG $\mathcal{G}_1$, there exists a pure PME 
$(\boldsymbol{\mathbbm{1}}_{\boldsymbol{a}^{*}_{i}},\boldsymbol{\mathbbm{1}}_{\boldsymbol{a}^{*}_{-i}})$ 
that can be computed in polynomial time by converting it into solving a pure ME of an auxiliary MPG $\mathcal{G}_2$.
\end{theorem}

{\it Proof:}
We construct an auxiliary MG $\mathcal{G}_2$, where the public state $e_N^t$ serves as the game state, the same transition probability $\hat{q}(\cdot)$ in \eqref{hat q} specifies the transition dynamics, each player's action is electricity demand $d^t_i$, and the stage payoff $g_i(\cdot)$ in \eqref{max_g} serves as the stage payoff:
\begin{align}\label{G2}
   \mathcal{G}_2 = \left\{\mathcal{T},\mathcal{N},\{\mathcal E_{N}^t\}_{t=1}^T,\left\{\mathcal{D}_{i}\right\}_{i=1}^{n},\left\{{g}_{i}\right\}_{i=1}^{n}, \hat q \right\}. 
\end{align} 

Accordingly, we represent a pure Markovian strategy of user $i$ in MG $\mathcal{G}_2$ by 
$\boldsymbol{\mathbbm{1}}_{\boldsymbol{d}_i} \in \mathcal{P}_i^{MS}$, 
$\boldsymbol{\mathbbm{1}}_{\boldsymbol{d}_i}(\cdot) := (\mathbbm{1}_{d_i^1}(\cdot), \ldots, \mathbbm{1}_{d_i^T}(\cdot))$ 
with $\mathbbm{1}_{d_i^t}(e_N^t) \in \mathcal{D}_i$ for all $t \in \mathcal{T}, e_N^t \in \mathcal{E}_N^t$. 
The value function under a pure strategy profile $(\boldsymbol{\mathbbm{1}}_{\boldsymbol{d}_i}, \boldsymbol{\mathbbm{1}}_{\boldsymbol{d}_{-i}})$ starting from $e_N^1$ is:
\begin{multline}\label{G3_value}
      \hat V^{\boldsymbol{\mathbbm{1}}_{\boldsymbol{d}_{i}},\boldsymbol{\mathbbm{1}}_{\boldsymbol{d}_{-i}}}_{1,i}(e_N^1) =  g_i(e^1_{N}, \boldsymbol{d}^{1}) +  \mathbf p^{1}_{2} \mathbf g_i^2(\boldsymbol{d}^{2}) + \mathbf p^{1}_{2}\mathbf Q ^{2}_{3}\mathbf g_i^3(\boldsymbol{d}^{3})  \\
      + \cdots + \mathbf p^{1}_{2}\mathbf Q^{2}_{3}\dots \mathbf Q^{T-1}_{T}\mathbf g_i^T(\boldsymbol{d}^{T}), ~ \forall e_N^1 \in \mathcal{E}_N^1,
\end{multline}
which clearly follows the structure of \eqref{eq:value_function}. Moreover, the constructed MG $\mathcal{G}_2$ satisfies the following two conditions:
\begin{enumerate}
    \renewcommand{\labelenumi}{\arabic{enumi})}
    \item The single-stage game is a potential game according to Lemma~\ref{one-horizon}, and
    \item The state transition probabilities are independent of users' actions.
\end{enumerate}
These two conditions are sufficient conditions in \cite{zhang2024gradient},\cite{leonardos2022global} for an MG to be an MPG. To make this explicit, we construct the potential function of $\mathcal{G}_2$ as:
\begin{multline}\label{eq7}
\phi_{e_N^1}\left(\boldsymbol{\mathbbm{1}}_{\boldsymbol{d}_{i}}, \boldsymbol{\mathbbm{1}}_{\boldsymbol{d}_{-i}}\right) = \phi(e^1_{N}, \boldsymbol{d}^{1}) +  \mathbf p^{1}_{2} \bm \phi^2(\boldsymbol{d}^{2}) + \mathbf p^{1}_{2}\mathbf Q ^{2}_{3}\bm \phi^3(\boldsymbol{d}^{3})  \\
      +  \cdots  + \mathbf p^{1}_{2}\mathbf Q^{2}_{3}\dots \mathbf Q^{T-1}_{T}\bm \phi^T(\boldsymbol{d}^{T}), ~ \forall e_N^1 \in \mathcal{E}_N^1,
\end{multline}
where the stage-wise potential function at state \( e^t_{N} \) is:
\begin{multline}\label{phi}
\phi(e^t_{N}, \boldsymbol{d}^{t}) =  \sum_{i=1}^{n} (\theta_i-\frac{\beta} {e^t_{N}+\gamma_{2}}) d_{i}^t-\frac{\alpha}{n e^t_{N} +\gamma_{1}} \sum_{i=1}^{n} (d_{i}^t)^2 \\
-\frac{\alpha}{n e^t_{N}+\gamma_{1}} \sum_{1 \leq i<j \leq n} d_{i}^t  d_{j}^t, ~ \forall e^t_{N} \in \mathcal E^t_{N},
\end{multline} 
and $\bm \phi^t(\boldsymbol{d}^{t}) = \big[ \phi(e^t_{N}, \boldsymbol{d}^{t}) \big]_{e^t_{N} \in \mathcal{E}^{t}_{N}}$ forms a column vector of values of the potential function across all states at time $t$.

It is easy to verify that the potential function \eqref{eq7} satisfies the condition in \eqref{Stochastic potential game function 1}, and thus the MG $\mathcal{G}_2$ is indeed an MPG.

Let $(\boldsymbol{\mathbbm{1}}_{\boldsymbol{d}^{*}_{i}},\boldsymbol{\mathbbm{1}}_{\boldsymbol{d}^{*}_{-i}})$ be a pure ME of $\mathcal{G}_2$, where $\mathbbm{1}_{d_i^{*t}}(e_N^t) = d^{*t}_i(e^{t}_{N}), \forall i \in \mathcal{N}, t \in \mathcal{T}, e_N^t \in \mathcal{E}_N$. Based on this, we construct a pure PMS profile in $\mathcal{G}_1$ by:
$(\boldsymbol{\mathbbm{1}}_{\boldsymbol{a}^{*}_{i}},\boldsymbol{\mathbbm{1}}_{\boldsymbol{a}^{*}_{-i}})$ where $\mathbbm{1}_{a_i^{*t}}(\boldsymbol s^t_i) = (d^{*t}_i(\boldsymbol s_i^t), c_i^{*t}(\boldsymbol s_i^t))
= (d^{*t}_i(e^{t}_{N}), d^{*t}_i(e^{t}_{N}) + b_i^t), \forall i \in \mathcal{N}, t \in \mathcal{T}, \boldsymbol s_i^t \in \mathcal{S}_i^t$. This PMS profile satisfies the structural constraints in \eqref{case2}, and its induced value function coincides with the equilibrium payoffs \eqref{eq:value_function} for the original game $\mathcal{G}_1$, thereby establishing a pure PME in $\mathcal{G}_1$.

Since $\mathcal{G}_2$ is an MPG with an explicit potential function, its pure equilibria can be obtained by maximizing the potential function. The potential optimization can be solved in polynomial time \cite{zhang2024gradient, leonardos2022global, shapley1953stochastic}, thereby implying that the PME of $\mathcal{G}_1$ is polynomial-time computable.
\te

In summary, Theorem~\ref{main thm} establishes the existence of a pure PME in $\mathcal{G}_1$, while Theorem~\ref{thm:MPG_equivalence} shows that this equilibrium is polynomial-time computable via the pure ME of the auxiliary MPG $\mathcal{G}_2$. This conversion reveals the underlying potential structure and enables efficient computation of equilibrium strategies. In the following section, we leverage this insight to design scalable algorithms for equilibrium computation and validate their performance through simulations.




\section{Algorithms and Simulations}\label{algsection}
\setcounter{table}{0} 
To solve the formulated Stackelberg Markov Game (SMG) for the smart grid presented in the above sections, we first compute the equilibrium of the lower-level Markov game, which is a key step for determining the overall pricing strategy.

Subsection~\ref{4.1} presents two algorithms for computing the PME of the lower-level MG: a centralized algorithm based on the potential function of MPG, which enables \textit{scalable computation} for large-scale systems, and a decentralized algorithm designed for scenarios with incomplete information. After establishing the lower-level equilibrium, Subsection~\ref{4.2} develops an algorithm for computing the optimal real-time pricing (RTP) for the upper level. Subsection~\ref{4.3} further extends the model to account for nonlinear user utility in electricity consumption.

\subsection{Equilibrium Computation in the Lower-Level Markov Game}\label{4.1}
\subsubsection{Centralized Algorithm via Finite Improvement Property (FIP) to Solve MPG \texorpdfstring{$\mathcal{G}_2$}{G2}}
By Theorem~\ref{thm:MPG_equivalence}, we only need to compute the ME of MPG $\mathcal{G}_2$ in order to solve the PME of MG $\mathcal{G}_1$.
Starting from any initial state $\boldsymbol s^1 \in \mathcal{S}^1$, our aim is to solve: \begin{align*} (\boldsymbol{\mathbbm{1}}_{\boldsymbol{d}^{\ast}_{i}}, \boldsymbol{\mathbbm{1}}_{\boldsymbol{d}^{\ast}_{-i}}) = \arg\max \limits_{\boldsymbol{\mathbbm{1}}_{\boldsymbol{d}_{i}}, \boldsymbol{\mathbbm{1}}_{\boldsymbol{d}_{-i}} } \phi_{\boldsymbol s^1}\left(\boldsymbol{\mathbbm{1}}_{\boldsymbol{d}_{i}}, \boldsymbol{\mathbbm{1}}_{\boldsymbol{d}_{-i}}\right), \end{align*}
where $\phi_{\boldsymbol s^1}(\cdot)$ is the potential function defined in \eqref{eq7}. 

\textit{i) Iterative Best-Response Process:} Below, we propose an iterative best-response process to compute this ME based on potential function $\phi_{\boldsymbol s^1}(\cdot)$. 

At iteration $k$, each user $i$ holds a pure Markovian strategy $\boldsymbol{\mathbbm{1}}_{\boldsymbol{d}_i^k}$, forming the strategy profile $(\boldsymbol{\mathbbm{1}}_{\boldsymbol{d}_i^k}, \boldsymbol{\mathbbm{1}}_{\boldsymbol{d}_{-i}^k})$.

\textit{Step 1 (Best response).} Given the opponents' strategy profile $\boldsymbol{\mathbbm{1}}_{\boldsymbol{d}^k_{-i}}$, each player $i \in \mathcal{N}$ computes its best response $\boldsymbol{\mathbbm{1}}_{\boldsymbol{\hat d}^k_{i}}$, i.e.,  \begin{align}\label{BR}
\boldsymbol{\mathbbm{1}}_{\boldsymbol{\hat d}^k_{i}} &=  \arg \max \limits_{\boldsymbol{\mathbbm{1}}_{\boldsymbol{d}_{i}^k} } \hat V^{\boldsymbol{\mathbbm{1}}_{\boldsymbol{d}^k_{i}}, \boldsymbol{\mathbbm{1}}_{\boldsymbol{d}^k_{-i}}}_{1,i}(\boldsymbol s^1) = \arg \max \limits_{\boldsymbol{\mathbbm{1}}_{\boldsymbol{d}^k_{i}} } \phi_{\boldsymbol s^1}\left(\boldsymbol{\mathbbm{1}}_{\boldsymbol{d}^k_{i}}, \boldsymbol{\mathbbm{1}}_{\boldsymbol{d}^k_{-i}}\right). 
\end{align}  

\textit{Step 2 (Payoff improvement).} Evaluate the corresponding payoff improvement for each player $i$: 
\begin{align}\label{delta}
\Delta V_{\boldsymbol s^1,i}^k = \hat V^{\boldsymbol{\mathbbm{1}}_{\boldsymbol{\hat d}^k_{i}}, \boldsymbol{\mathbbm{1}}_{\boldsymbol{d}^k_{-i}}}_{1,i}(\boldsymbol s^1) -\hat V^{\boldsymbol{\mathbbm{1}}_{\boldsymbol{d}^k_{i}}, \boldsymbol{\mathbbm{1}}_{\boldsymbol{d}^k_{-i}}}_{1,i}(\boldsymbol s^1). \end{align} 

\textit{Step 3 (Strategy update).} Choose the player with the largest improvement, i.e., $j=\arg\max_{i \in \mathcal{N}}$ $\Delta V_{\boldsymbol s^1,i}^k$, and update its strategy, while others remain unchanged: \begin{align*} \boldsymbol{\mathbbm{1}}_{\boldsymbol{d}^{k+1}_{j}} \leftarrow \boldsymbol{\mathbbm{1}}_{\boldsymbol{\hat d}^k_{j}}, \quad \boldsymbol{\mathbbm{1}}_{\boldsymbol{d}^{k+1}_{i}} \leftarrow \boldsymbol{\mathbbm{1}}_{\boldsymbol{d}^k_{i}}, ~ \forall i \neq j. \end{align*} 

Repeat Steps 1 to 3 until no player can improve its payoff, i.e., $\Delta V_{\boldsymbol s^1,i}^k = 0, \ \forall i \in \mathcal{N}$. 
 
The potential function $\phi_{\boldsymbol s^1}(\cdot)$ strictly increases with each update in Steps 2 and 3. Since the game $\mathcal{G}_2$ is a finite potential game, the above process satisfies the Finite Improvement Property (FIP) \cite{monderer1996potential}, thus the process converges to a pure ME in finite steps \cite{chew2016potential}. The complete procedure is summarized in Algorithm \ref{FIP}.

\begin{algorithm}[H]
\caption{Centralized Algorithm via FIP to Solve MPG $\mathcal{G}_2$.}
\label{FIP}
\begin{algorithmic}
\STATE \textbf{Input:} $K_{\text{max}},\mathcal{T}, \mathcal{N}, \{\mathcal{E}_N^t\}_{t=1}^T,\left\{\mathcal{D}_{i}\right\}_{i=1}^{n}, \{\mathcal{A}_i\}_{i=1}^{n}, \{V_i\}_{i=1}^{n}$.
\STATE \textbf{Initialization:}
\STATE \hspace{0.5cm} Randomly initialize profile $(\boldsymbol{\mathbbm{1}}_{\boldsymbol{d}^1_{i}}, \boldsymbol{\mathbbm{1}}_{\boldsymbol{d}^1_{-i}})$.
\STATE \textbf{For} $k = 1$ to $K_{\text{max}}$
\STATE \hspace{0.5cm} \textbf{for each} $i \in \mathcal{N}$ \textbf{do}
\STATE \hspace{1.0cm} Compute best response $\boldsymbol{\mathbbm{1}}_{\boldsymbol{\hat d}^k_i}$ to $\boldsymbol{\mathbbm{1}}_{\boldsymbol{d}^k_{-i}}$. 
\STATE \hspace{1.0cm} Evaluate payoff improvement: $\Delta V_{\boldsymbol s^1,i}^k$ in \eqref{delta}.
\STATE \hspace{0.5cm} \textbf{end for}
\STATE \hspace{0.5cm} Identify $j = \arg\max_{i \in \mathcal{N}} \Delta V_{\boldsymbol s^1,i}^k$.
\STATE \hspace{0.5cm} \textbf{if} $\Delta V_{\boldsymbol s^1,j}^k > 0$ \textbf{then}
\STATE \hspace{1.0cm} Update: $\boldsymbol{\mathbbm{1}}_{\boldsymbol{d}^{k+1}_{j}} \gets \boldsymbol{\mathbbm{1}}_{\boldsymbol{\hat d}^k_j}$, $\boldsymbol{\mathbbm{1}}_{\boldsymbol{d}^{k+1}_i} \gets \boldsymbol{\mathbbm{1}}_{\boldsymbol{d}^k_i}, \forall i \neq j$.
\STATE \hspace{0.5cm} \textbf{else if} $\Delta V_{\boldsymbol s^1,j}^k = 0$ \textbf{then}
\STATE \hspace{1.0cm} Convergence achieved; stop iteration.
\STATE \hspace{0.5cm} \textbf{end if}
\STATE \hspace{0.5cm} Increment $k \gets k + 1$.
\STATE \textbf{end for}
\STATE \textbf{Output:} Converged pure ME: $ (\boldsymbol{\mathbbm{1}}_{\boldsymbol{d}^{\ast}_i}, \boldsymbol{\mathbbm{1}}_{\boldsymbol{d}^{\ast}_{-i}})$.
\end{algorithmic}
\end{algorithm}

\textit{ii) Accelerated Best-Response Computation:} In Algorithm~\ref{FIP}, the main step is solving the best response MS $\boldsymbol{\mathbbm{1}}_{\boldsymbol{\hat d}^k_{i}}$ \eqref{BR}, which is generally hard. However, by Theorem~\ref{thm:MPG_equivalence}, the potential function $\phi_{\boldsymbol s^1}(\cdot)$ is a linear combination of stage-wise potential functions $\phi(e^t_{N}, \boldsymbol{d}^t)$ (see \eqref{phi}). This allows us to decompose the problem in~\eqref{BR} into independent subproblems for each stage $t$ and state $e^{t}_{N}$:
\begin{align}\label{independent subproblems} 
\hat {d}_i^{t}(e^{t}_{N}) = \arg\max\limits_{{d}_i^{t}(e^{t}_{N})} \phi(e^{t}_{N}, \boldsymbol{d}^{t}),~ \forall t\in \mathcal{T},e^{t}_{N} \in \mathcal E^t_{N}, \end{align} 
with opponents' strategies $\boldsymbol{d}_{-i}^{t}$ fixed.
Furthermore, in \eqref{independent subproblems}, we only need to consider the term in $\phi(e^t_{N}, \boldsymbol{d}^t)$ that involves $d_i^t$, which is given by:
\begin{align}\label{f(d_i^t)}
f(d_i^t) = (\theta_i-\frac{\beta} {e^t_{N}+\gamma_{2}}-\frac{\alpha}{n e^t_{N}+\gamma_{1}}\sum_{j\neq i}d_j^t) d_{i}^t -\frac{\alpha}{n e^t_{N}+\gamma_{1}}  (d_{i}^t)^2.
\end{align}
Obviously, $f(d_i^t)$ is a strictly concave quadratic function in $d_i^t$.
Thus we can get the optimal (continuous) response:
\begin{align}\label{t,opt}
d_i^{t,\mathrm{opt}} = \left(\theta_i - \frac{\beta}{e^{t}_{N}+\gamma_2} - \frac{\alpha}{n e^{t}_{N} + \gamma_1} \sum_{j \neq i} d_j^t \right)\frac{n e^{t}_{N} + \gamma_1}{\alpha}. 
\end{align} 
If it lies outside the discrete action set $\mathcal{D}_i$, the two nearest discrete values are compared to select the best.

The above idea is summarized in Subroutine~\ref{SubroutineFIP} below.

\setcounter{algorithm}{0}
{
\renewcommand{\thealgorithm}{Subroutine~\arabic{algorithm}} 
\makeatletter
\renewcommand{\fnum@algorithm}{\thealgorithm} 
\makeatother
\begin{algorithm}[H]
\caption{Accelerated BR Computation.}
\label{SubroutineFIP}
\begin{algorithmic}
\STATE \textbf{Input:} $\mathcal{D}_i, e^t_{N},{d}^t_{-i}$.
\STATE \textbf{Procedure:}
\STATE \hspace{0.5cm} 1. Fix $d^t_{-i}$ and extract the part of $\phi(e^t_{N}, \boldsymbol{d}^t)$ involving 
\STATE \hspace{1cm} $d_i^t$, i.e., $f(d_i^t)$ in \eqref{f(d_i^t)}.
\STATE \hspace{0.5cm} 2. Compute continuous optimum $d_i^{t,\mathrm{opt}}$ based on \eqref{t,opt}. 
\STATE \hspace{0.5cm} 3. Find two nearest actions $d_1, d_2 \in \mathcal{D}_i$ such that 
\STATE \hspace{1.0cm} $a_1 \leq d_i^{t,\mathrm{opt}} \leq a_2$.
\STATE \hspace{0.5cm} 4. Select best discrete action:
\STATE \hspace{1.0cm} $\hat{d}_i^t(e^t_{N}) = \arg\max_{d_i^t \in \{d_1, d_2\}} f(d_i^t)$.
\STATE \hspace{0.5cm} 5. Return $\hat{d}_i^t(e^t_{N})$.
\STATE \textbf{Output:} Best response $\hat{d}_i^t(e^t_{N})$.
\end{algorithmic}
\end{algorithm}
}

By applying the centralized Algorithm~\ref{FIP} with the embedded Subroutine~\ref{SubroutineFIP}, we can compute the ME of $\mathcal{G}_2$, from which the PME of $\mathcal{G}_1$ follows directly by Theorem~\ref{thm:MPG_equivalence}, completing the lower-level equilibrium computation. 

\textit{This method has two key advantages.} First, it guarantees convergence in finite steps. Second, the simplified state space $\mathcal{E}^t_{N}$ does not grow with the number of users, which makes the algorithm scalable to large systems. This scalability is crucial in power grid applications, yet rarely achieved in previous studies.

To demonstrate the effectiveness of Algorithm~\ref{FIP}, we conduct a simulation using real data with 50 users over 7 stages.

\begin{figure}[htb]
\centering
\includegraphics[width=0.95\textwidth, height=6cm]{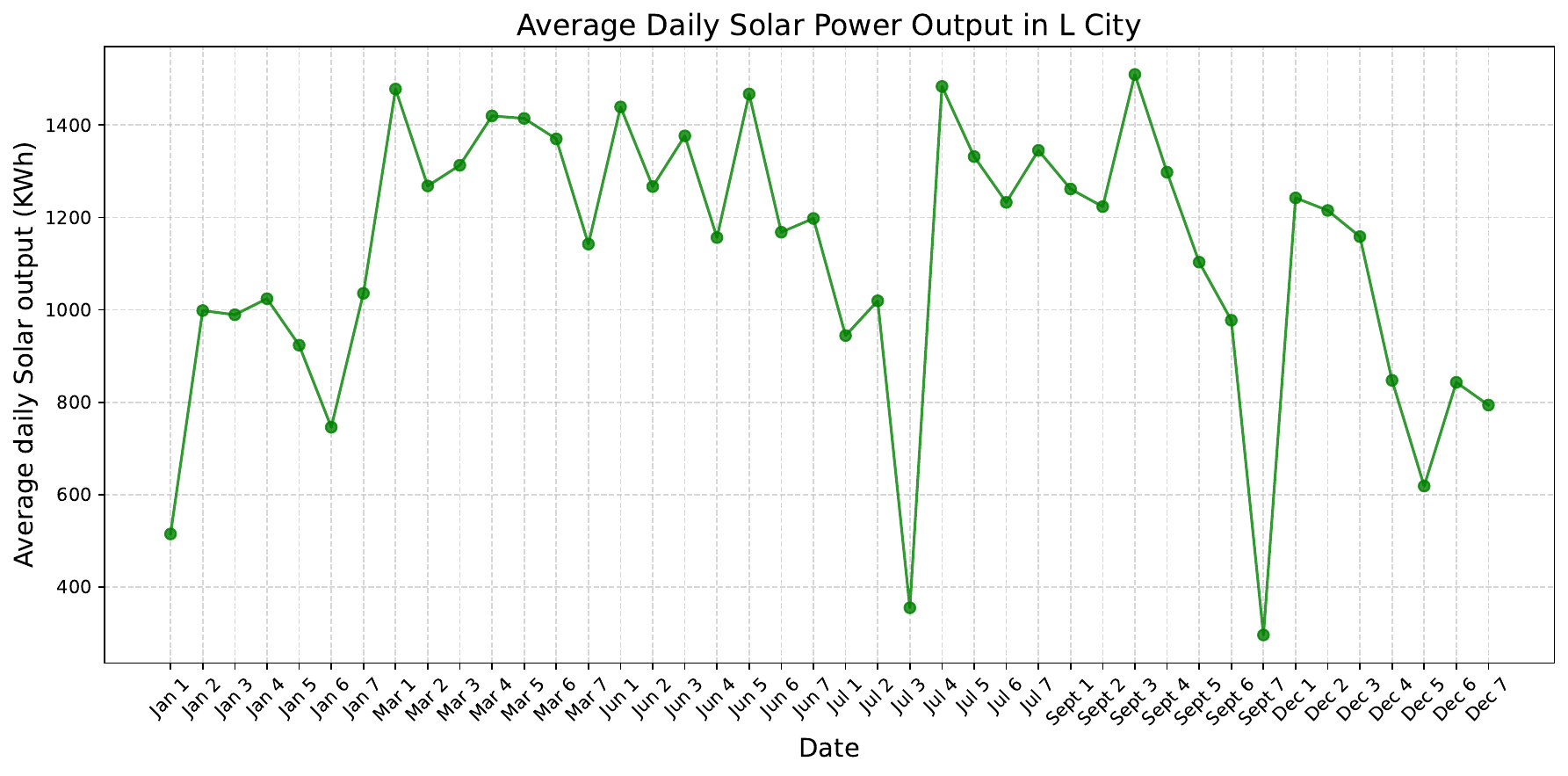}
\caption{Average hourly solar generation in City~L, computed from minute-level data collected during the first week of each month over a six-month period.}
\label{price picture}
\end{figure}

\textit{iii) Simulation:} Benefiting from these structural advantages, the proposed method enables efficient equilibrium computation in \textit{large-scale, multi-user, long-horizon MG settings}. To conduct simulations, we use real-world solar generation data from a region in a Chinese province.\footnote{The dataset was provided by China Southern Power Grid and has been anonymized for publication.}  Fig.~\ref{price picture} shows the average hourly generation in a day during the first week of each month over a six-month period in City~L.

To approximate a 50-user system, we scale the original generation values by  0.1, reflecting a suitable per-user electricity demand~\cite{liang2022impacts}, and denote the processed value on day~$j$ of month~$i$ as $\zeta_{ij}$, $i = 1,\dots,6$ and $j = 1,\dots,7$. The predicted uncontrollable generation for day~$j$ is then defined as the monthly average: $\hat{e}_N^j := \frac{1}{6} \sum_{i=1}^6 \zeta_{ij}, \quad j = 1,\dots,7$, yielding the predicted uncontrollable electricity sequence: $\{\hat{e}_N^t\}_{t=1}^7 = [50, 110, 90, 130, 80, 70, 100]$.

We compute deviations from the predicted values as $\widetilde{\zeta}_{ij} := \zeta_{ij} - \hat{e}_N^j$, and discretize them to $\{-20, 0, 20\}$, which serve as possible realizations of the forecast error $\omega_N^t \in \{-20, 0, 20\}$. The transition probabilities of $\omega_N^t$ are estimated by empirical frequency. For instance, the probability of transitioning from $\omega_N^{t} =20$ to $\omega_N^{t+1} =-20$ is given by:
\begin{align*}
\hat{q}(\omega_N^{t+1} = -20 \mid \omega_N^t = 20) = \frac{\sum_{i=1}^6 \sum_{j=1}^6 \mathbb{I}_{\{ \widetilde{\zeta}_{ij} = 20, \widetilde{\zeta}_{i,j+1} = -20 \}}}{\sum_{i=1}^6 \sum_{j=1}^7 \mathbb{I}_{\{ \widetilde{\zeta}_{ij} = 20 \}}}, ~\forall t = 1,\dots,6,
\end{align*}
where $\mathbb{I}_{\{\cdot\}}$ is the indicator function. Other transition probabilities are computed similarly, yielding the transition matrix:
\begin{align}\label{example hat q}
\hat{q}(\omega^{t+1}_{N} \mid \omega^t_{N}) =
\begin{bmatrix}
5/11 & 5/11 & 1/11 \\
1/4 & 7/16 & 5/16 \\
2/9 & 4/9 & 1/3
\end{bmatrix},\forall t=1,\dots,6,
\end{align}
where $\omega^t_{N}$ and $\omega^{t+1}_{N}$ take values in $\{20,0,-20\}$.

\begin{example}[50-User Equilibrium Computation in MPG]\label{FIP example}
Now, we consider a 50-user, 7-horizon MPG defined by \eqref{G2}, where $\mathcal{T} = \{1,\dots,7\}$ and $\mathcal{N} = \{1,\dots,50\}$. The state space is $\{\mathcal{E}_N^t\}_{t=1}^{T=7}$, where $\hat{e}^t_N$ and $\omega^t_N$ are obtained from the real-world data as described above. Each user's action space is $\mathcal{D}_i = \{0,1,2,3,4\}$. The reward function $r_i(e^t_{N}, \boldsymbol{d}^t)$ follows \eqref{current reward}, with $\theta_i$ drawn from the uniform distribution on $[0.9, 1.5]$, for $i=1,\dots,50$ (see Appendix~\ref{appendix4} for full list). The pricing parameters are set as $\alpha = 19$, $\beta = 20$, and $\gamma_1 = \gamma_2 = 1$, ensuring the resulting electricity price remains within a realistic range of approximately 0.5 to 1.5 CNY/kWh~\cite{lin2019does}.

A pure PMS for user $i$ is represented as $\boldsymbol{\mathbbm{1}}_{\boldsymbol{d}_i} =[ e^1, e^2,$ $\dots, e^{21}],$ 
over the $|\mathcal{T}| \cdot |\mathcal{E}_{N}^t| = 21$ state-time pairs, where each $e^j$ is a one-hot vector of dimension 5 indicating the selected action under the $j$-th state pair. The complete strategy profile of all 50 users forms a $50 \times 21 \times 5$ tensor.
We use $\text{strategy}[i][j][m]$ to denote whether user $i$ selects action $m$ under the $j$-th state pair (0 or 1 for pure strategies).

Using Algorithm \ref{FIP}, the results shown in Fig.~\ref{FIP picture} demonstrate that the FIP-based algorithm quickly converges to a pure ME, and the best-response deviations $\Delta V_{\boldsymbol{s}^1,i}^k$ in \eqref{delta} vanish after a finite number of steps, with the top black curve representing $\max_{i \in \mathcal{N}} \Delta V_{\boldsymbol{s}^1,i}^k$, confirming the algorithm's validity.

\begin{figure}[t]
    \centering
    \captionsetup[subfigure]{labelformat=empty}

    \includegraphics[width=0.85\columnwidth, keepaspectratio,
        trim=2mm 1mm 1mm 0mm, clip]{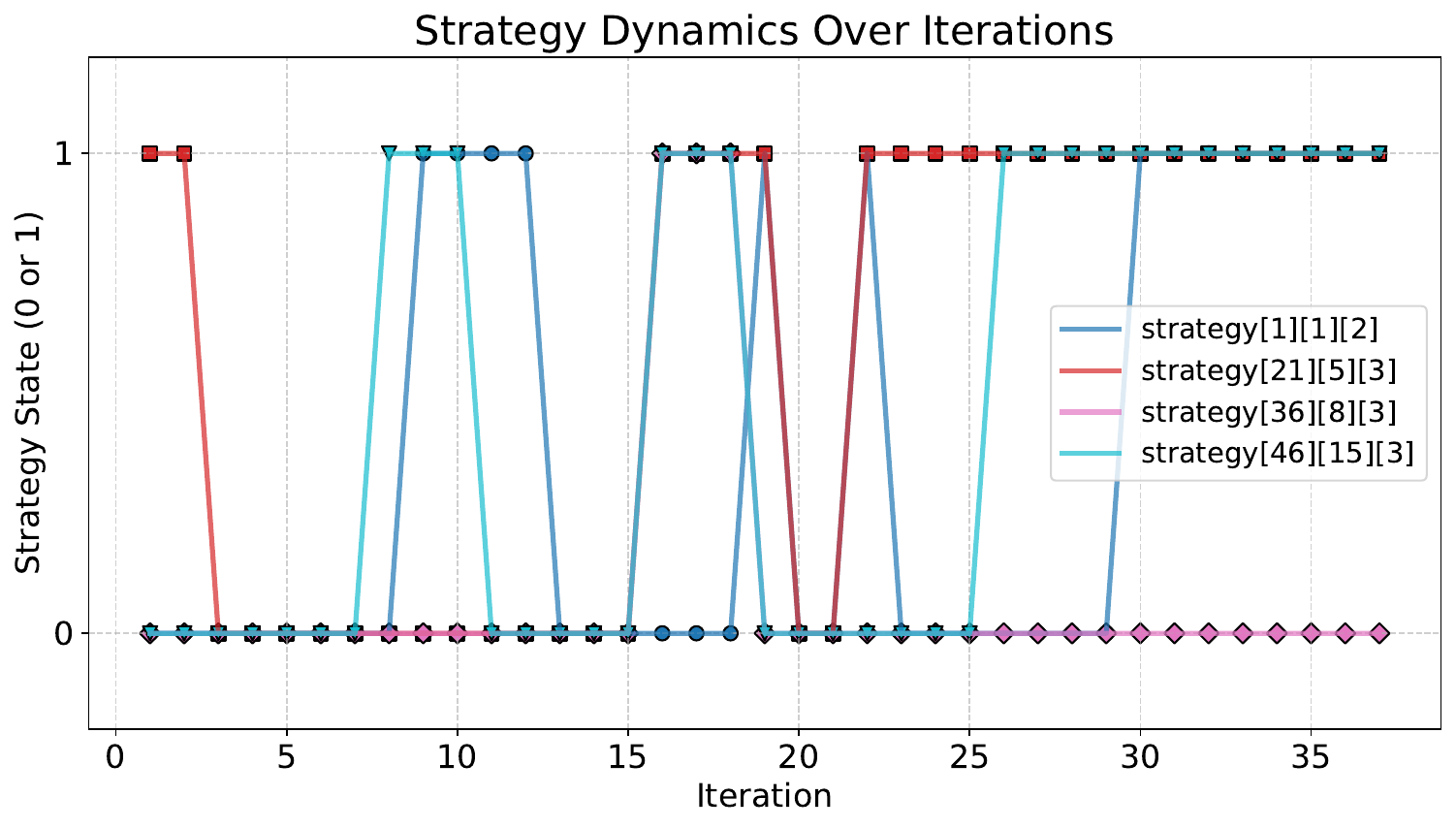}

    \vspace{0.2cm}  

    \includegraphics[width=0.85\columnwidth, keepaspectratio,
        trim=2mm 1mm 1mm 0mm, clip]{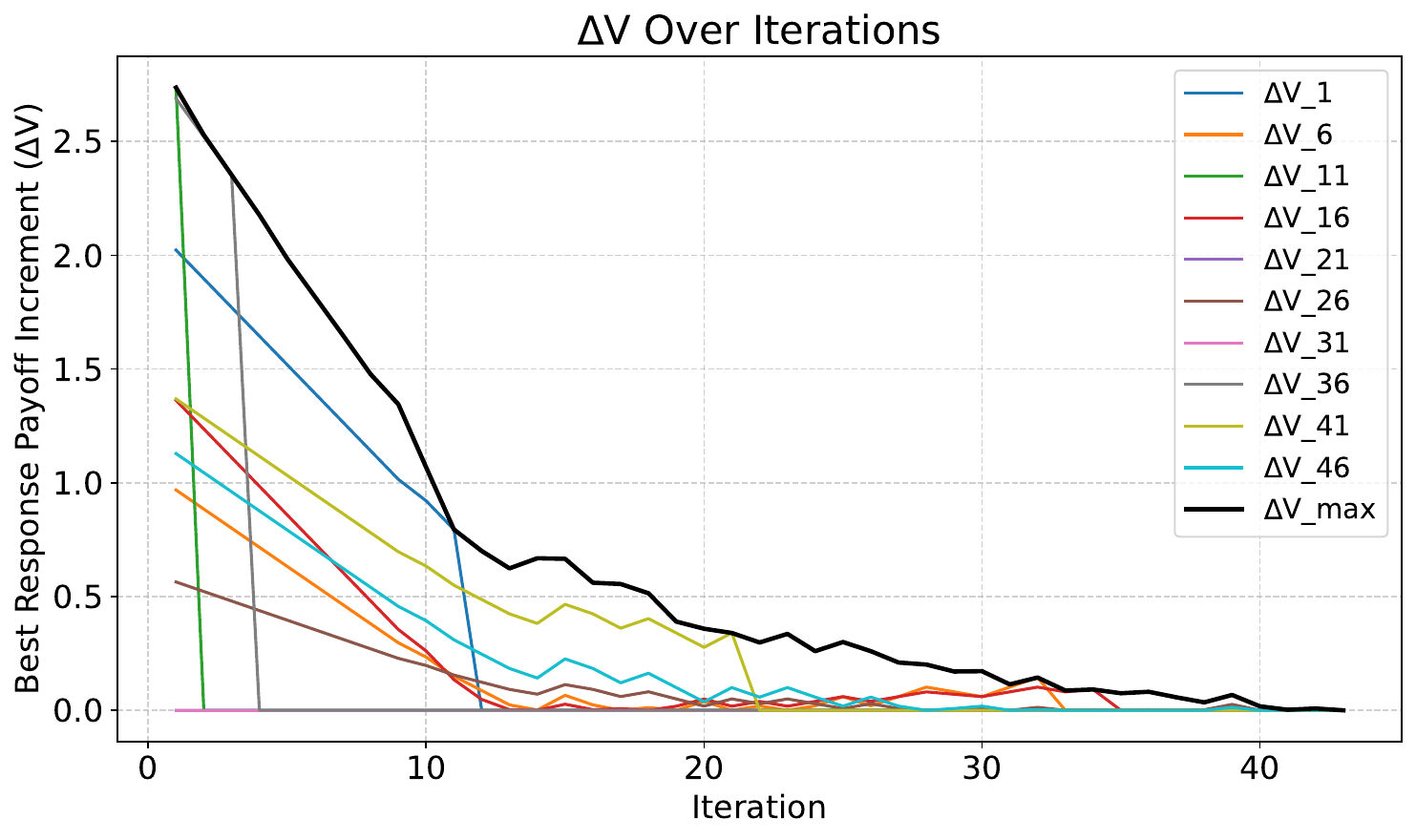}

    \caption{FIP algorithm converges to a pure ME in Example~\ref{FIP example}.}
    \label{FIP picture}
\end{figure}

In Appendix~\ref{appendix4}, we present the pure equilibrium demand of the 50 users under $e_{N}^1 = 70$ and $e_{N}^3 = 90$. Fig.~\ref{Q picture} illustrates that users’ linear electricity consumption coefficients $\theta_i$ generally increase the equilibrium demand, which aligns with intuition. In high renewable scenarios (e.g., $e_{N}^3 = 90$), some users with higher $\theta_i$ purchase less, possibly due to multiple equilibria.

\begin{figure}[htb]
    \centering
    \includegraphics[width=0.85\columnwidth]
    {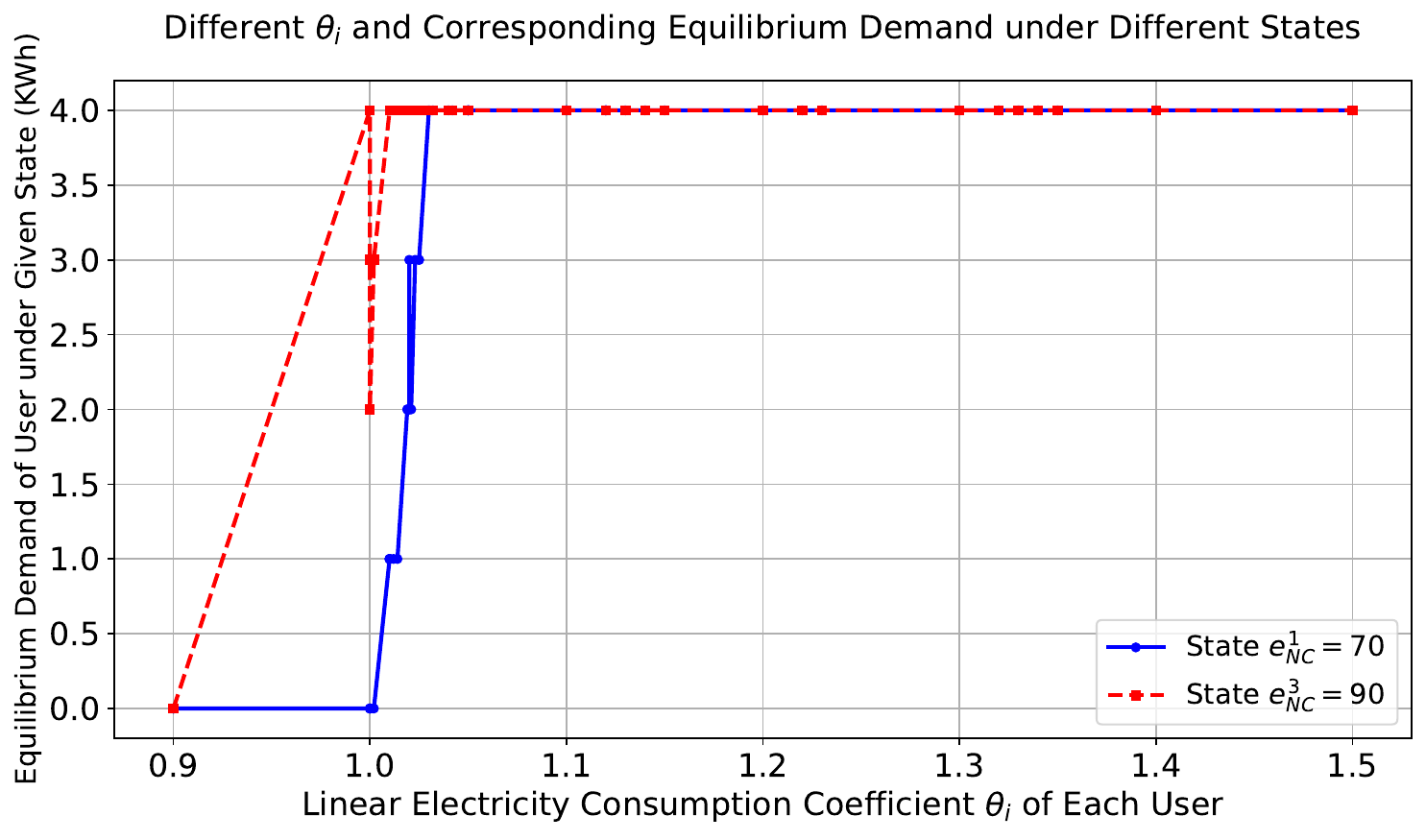}  
    \caption{Different $\theta_i$ and corresponding equilibrium demand.}
    \label{Q picture}
\end{figure}

\end{example}

\subsubsection{Decentralized Algorithm via FP + MDP to Solve MG \texorpdfstring{$\mathcal{G}_1$} {G1}}\label{second alg}

Algorithm~\ref{FIP} relies on complete information about all users' strategies and payoffs. Here, we present a decentralized learning algorithm, FP + MDP, suitable for incomplete information environments.

In the $k$-th iteration, each user $i$ maintains its PMS $\boldsymbol \pi_i^k: \cup_{t=1}^{T}\{ t \} \times \mathcal{S}_i^t \rightarrow \cup_{t=1}^{T}\Delta(\mathcal{A}_i^t)$ \footnote{Here strategy $\boldsymbol \pi_i^k: \cup_{t=1}^{T}\{ t \} \times \mathcal{S}_i^t \rightarrow \cup_{t=1}^{T}\Delta(\mathcal{A}_i^t)$ is a compact form of the stage-wise representation in Definition~\ref{PMS}, used for algorithmic updates.} and an estimated strategy of others, denoted by $\hat{\boldsymbol \pi}_{-i}^k: \cup_{t=1}^{T} \{ t \} \times \mathcal{E}_N^t \rightarrow \cup_{t=1}^{T}\Delta(\mathcal{D}^t_{-i})$, where $\mathcal{D}^t_{-i} := \{d_{-i}^t \mid d_{-i}^t = \sum_{j \neq i} d_j^t, \ \forall t \in \mathcal{T} \}$.

It is reasonable for each user $i$ to track only $\hat{\boldsymbol \pi}_{-i}^{k}$ for two reasons. First, the stage payoff of user $i$ depends on others solely through their aggregate demand $ d_{-i}^t$. Second, by Theorem~\ref{main thm}, the equilibrium demand $d_i^{*t}$ is independent of the storage level $b_i^t$. Hence, the estimated strategy $\hat{\boldsymbol \pi}_{-i}^k$ can be defined over $ \cup_{t=1}^{T}\{ t \} \times \mathcal{E}_N^t$.

Then the iterative process proceeds as follows:

\textit{Step 1 (Best-response PMS).}
Given the estimated strategy $\hat{\boldsymbol \pi}_{-i}^{k}$, user $i$ solves the best-response PMS $\tilde{\boldsymbol \pi}_{i}^{k}$ by solving a corresponding Markov Decision Process (MDP) $M_i^k$ , which can be solved via backward induction. See Appendix~\ref{appendix5} for the MDP model and Algorithm~\ref{MDP}.

\textit{Step 2 (Data collection).} 
All users play the MG $\mathcal{G}_1$ using the best-response PMS profile $\tilde{\boldsymbol \pi}^k = (\tilde{\boldsymbol \pi}_{1}^{k}, \dots, \tilde{\boldsymbol \pi}_{n}^{k})$. During this episode, each user $i$ records the sequence of visited public states $\tilde e_N^t$ and the corresponding aggregate demands of others $ d_{-i}^{t}$ (inferred from electricity prices):
\begin{align}\label{sequence}
(1,\tilde e_{N}^{1},  d_{-i}^{1}; \; 2,\tilde e_{N}^{2},  d_{-i}^{2}; \; \ldots; \; T, \tilde e_{N}^{T}, d_{-i}^{T})_{-i}^{k}.  \end{align} 

\textit{Step 3 (PMS and estimated strategy updates).} 
Each user updates its own policy $\boldsymbol \pi_i^k$ and estimated strategy $\hat{\boldsymbol \pi}_{-i}^k$ following a Fictitious Play-style update. Different from that in normal-form games, $\hat{\boldsymbol \pi}_{-i}^{k}$ is updated only at visited states $\tilde e_N^t$ in~\eqref{sequence}, and remains unchanged elsewhere.

\textit{PMS Update:}
\begin{multline}\label{i strategy update}
\boldsymbol \pi_{i}^{k+1}(\cdot \mid t,e_{N}^{t},b_{i}^{t})  = 
(1-\alpha_{\#}^{k+1})\boldsymbol \pi_{i}^{k}(\cdot \mid t,e_{N}^{t},b_{i}^{t})
+ \alpha_{\#}^{k+1}  \tilde{\boldsymbol \pi}^k (\cdot \mid t,e_{N}^{t},b_{i}^{t}),  ~\forall (t,e_{N}^{t},b_{i}^{t}) \in \mathcal{T} \times \mathcal S_{i}.
\end{multline}

\textit{Estimated Strategy Update:}
\begin{align}\label{i estimate update}
\hat{\boldsymbol \pi}_{-i}^{k+1}(\cdot \mid t,e_{N}^{t})  
\quad\quad\quad\quad\quad\quad\quad\quad\quad\quad\quad\quad\quad\quad\quad\quad\quad\quad\quad\quad\quad\quad\quad\quad  \notag \\
= \begin{cases}
\begin{aligned}
(1-\beta_{\#}^{k+1})\hat{\boldsymbol \pi}_{-i}^{k}(\cdot \mid t,e_{N}^{t}) + \beta_{\#}^{k+1}\mathbf{1}_{d_{-i}^{t}},~ &\quad\text{if } (t, e_{N}^{t})~ \text{appears in}~ \eqref{sequence},
\end{aligned} \\
\hat{\boldsymbol \pi}_{-i}^{k}(\cdot \mid t,e_{N}^{t}),  \quad\quad\quad\quad\quad\quad\quad\quad\quad\quad~~\text{otherwise}.
\end{cases}
\end{align} 
where $\mathbf{1}_{d_{-i}^{t}} \in \Delta(\mathcal{D}_{-i})$ denotes the pure strategy that assigns probability one to the aggregate demand $d_{-i}^t$ corresponding to $(t, e_N^t)$ in \eqref{sequence}. In \eqref{i estimate update}, the step sizes $\alpha_{\#}^{k}$ and $\beta_{\#}^{k}$ satisfy standard diminishing conditions:
\begin{align*}
\sum_{k=1}^\infty \alpha_{\#}^{k}= \infty, \quad \sum_{k=1}^\infty (\alpha_{\#}^{k})^2 < \infty, \notag ~~~
\sum_{k=1}^\infty \beta_{\#}^{k}= \infty, \quad \sum_{k=1}^\infty (\beta_{\#}^{k})^2 < \infty,
\end{align*}
see, e.g., \cite{sayin2021decentralized, mertikopoulos2024unified}. In our experiment, $\alpha_{\#}^{k}= \beta_{\#}^{k}= \frac{1}{k}$.

The complete procedure is detailed in Algorithm~\ref{FP+MDP}.

\begin{algorithm}[H]
\caption{Decentralized Algorithm via FP + MDP to Solve MG $\mathcal{G}_1$.}
\label{FP+MDP}
\begin{algorithmic}
\STATE \textbf{Input:} $\mathcal{T}, \mathcal{N}, \{\mathcal{S}_i^t\}_{t=1}^T, \{\mathcal{A}_i\}_{i=1}^n, \{r_i\}_{i=1}^n, q, K$, and step sizes $\alpha_{\#}^k$, $\beta_{\#}^k$.
\STATE \textbf{Initialization:}
\STATE \hspace{0.5cm} Initialize $\boldsymbol \pi_i^1$ and $\hat{\boldsymbol \pi}_{-i}^1$ as uniform policies, $\forall i \in \mathcal{N}$.
\STATE \textbf{For} $k = 1, 2, \dots, K$ \textbf{do}
\STATE \hspace{0.5cm} \textbf{For each} $i \in \mathcal{N}$ \textbf{do}
\STATE \hspace{1.0cm} Construct MDP $M_i^k := (\mathcal{T}, \{\mathcal{S}_i^t\}_{t=1}^T, \mathcal{A}_i, q_i^k, r_i^k)$ based on the estimation $\hat{\boldsymbol \pi}_{-i}^{k}$.
\STATE \hspace{1.0cm} Solve $M_i^k$ using Algorithm~\ref{MDP} to obtain BR $\tilde{\boldsymbol \pi}_i^{k}$.
\STATE \hspace{0.5cm} \textbf{end for}
\STATE \hspace{0.5cm} Form the strategy profile $\tilde{\boldsymbol \pi}^{k} = (\tilde{\boldsymbol \pi}_1^{k}, \dots, \tilde{\boldsymbol \pi}_n^{k})$, play the $T$-horizon MG under $\tilde{\boldsymbol \pi}^{k}$.
\STATE \hspace{0.5cm} \textbf{For each} $i \in \mathcal{N}$ \textbf{do}
\STATE \hspace{1.0cm} Observe trajectory $(1,\tilde e_{N}^{1}, d_{-i}^{1}; \dots; T, \tilde e_{N}^{T}, d_{-i}^{T})_{-i}^k$, generate the pure strategy $\mathbf{1}_{d_{-i}^{t}}$.
\STATE \hspace{1.0cm} \textbf{Policy Update:}
\STATE \hspace{1.2cm} Update $\boldsymbol \pi_i^{k+1}$ via~\eqref{i strategy update}.
\STATE \hspace{1.2cm} Update $\hat{\boldsymbol \pi}_{-i}^{k+1}$ via~\eqref{i estimate update}.
\STATE \hspace{0.5cm} \textbf{end for}
\STATE \textbf{end for}
\STATE \textbf{Output:} Final strategy profile $\boldsymbol \pi^K = (\boldsymbol \pi_1^K, \dots, \boldsymbol \pi_n^K)$.
\end{algorithmic}
\end{algorithm} 

Algorithm~\ref{FP+MDP} does not require the MG $\mathcal{G}_1$ to be a potential game. It offers a natural and practical decentralized learning approach for general nonzero-sum Markov games. Theoretical convergence analysis remains challenging and may require weaker equilibrium notions such as coarse correlated equilibrium (CCE) \cite{mao2023provably}, which we leave for future work.

We next provide numerical results to demonstrate the effectiveness of Algorithm~\ref{FP+MDP}.

\begin{example}[Decentralized Equilibrium Learning in MG]\label{FP+MDP example}
Consider a 3-user, 3-horizon MG defined by \eqref{G1}, where $\mathcal{T} = \{1,2,3\}$, $\mathcal{N} = \{1,2,3\}$, and the state space is $\{\mathcal{S}^t\}_{t=1}^{T=3}$ where $\mathcal{S}^t = \mathcal{E}_N^t \times \prod_{i=1}^{n} \mathcal{B}_{i}$ with the private component is $\mathcal{B}_i=\{0,1,2\},  \forall i \in \mathcal{N}$. At time $t$, the public component is $\mathcal{E}_N^t = \{e^t_N \mid e^t_N = \hat e^t_N + \omega, \omega \in \{2,0,-2\}\}$ with forecasts $\{\hat e^1_N,\hat e^2_N,\hat e^3_N\}=[5,11,8]$.
The transition probability $q(\cdot)$ satisfies \eqref{q}, and $\hat q (\cdot)$ follows \eqref{example hat q}, with $\omega^t_N$ restricted to $\{2,0,-2\}$. Each user's action space is $\mathcal{A}_i$ with $\mathcal{D}_i = \{0,1,2,3,4\}$ and $\mathcal{C}_i = \{0,1,2,3,4,5,6\}$. The reward function $r_i(\boldsymbol{s}^t,\boldsymbol{a}^t)$ follows \eqref{current reward}, where $\{\theta_i\}_{i=1}^3 = [0.9, 1, 1.1]$. The electricity pricing parameters are $\alpha = 1.5$, $\beta = 1.5$, and $\gamma_1 = \gamma_2 = 1$.

In Algorithm~\ref{FIP}, since the focus is on MPG, we use the best-response deviation $\Delta V_{\boldsymbol{s}^1,i}^k$ in \eqref{delta} to measure the distance between a player's current strategy and its equilibrium strategy.
In this example, we adopt the more general NashConv metric \cite{lanctot2017unified} to assess the closeness of the approximate equilibrium:
\begin{align*}
    NashConv &:=  \frac{1}{|\mathcal{S}^1|}\sum_{\boldsymbol s^{1} \in \mathcal{S}^1} \frac{1}{n}\sum_{i \in \mathcal{N}}(V^{\textbf{BR}_i(\boldsymbol \pi),\boldsymbol \pi_{-i}}_{1,i}(\boldsymbol s^{1})-V^{\boldsymbol \pi_i,\boldsymbol \pi_{-i}}_{1,i}(\boldsymbol s^{1})),
\end{align*} where NashConv $= 0$ indicates an equilibrium.

Fig.~\ref{FP+MDP picture} shows that all users’ strategies gradually stabilize, and the NashConv decreases toward zero, indicating that the algorithm effectively approximates the PME. Compared to Example~\ref{FIP example}, the required number of iterations increases by one to two orders of magnitude, reflecting the increased complexity of equilibrium computation in MGs and the trade-off introduced by decentralization.
\begin{figure}[t]
	\centering
	\captionsetup[subfigure]{labelformat=empty}

	\begin{subfigure}{0.492\linewidth} 
		\centering
		\includegraphics[width=\linewidth,trim=1.2cm 0.1cm 2.2cm 0.8cm, clip]{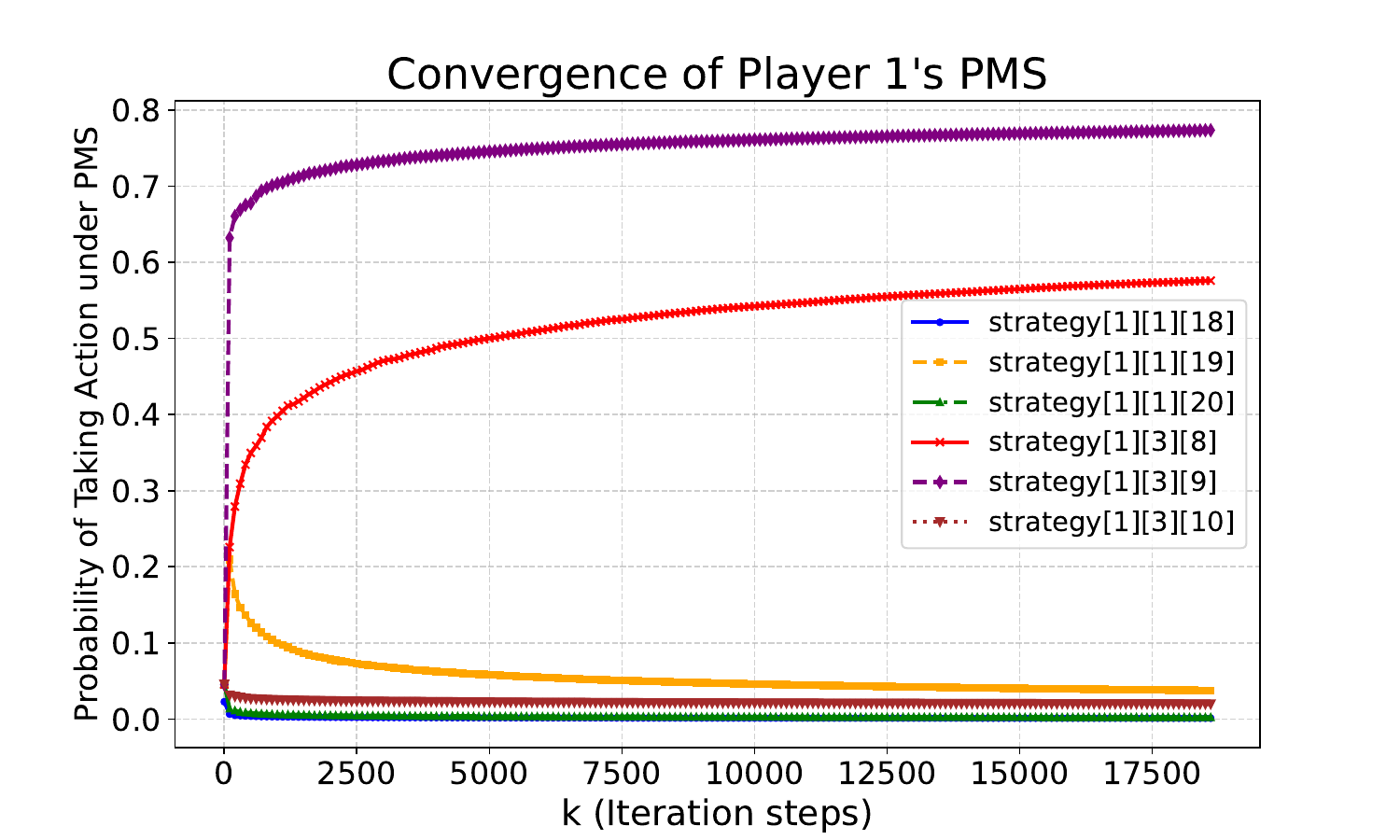}
	\end{subfigure}
	\hfill
	\begin{subfigure}{0.492\linewidth}
		\centering
		\includegraphics[width=\linewidth,trim=1.2cm 0.1cm 2.2cm 0.8cm, clip]{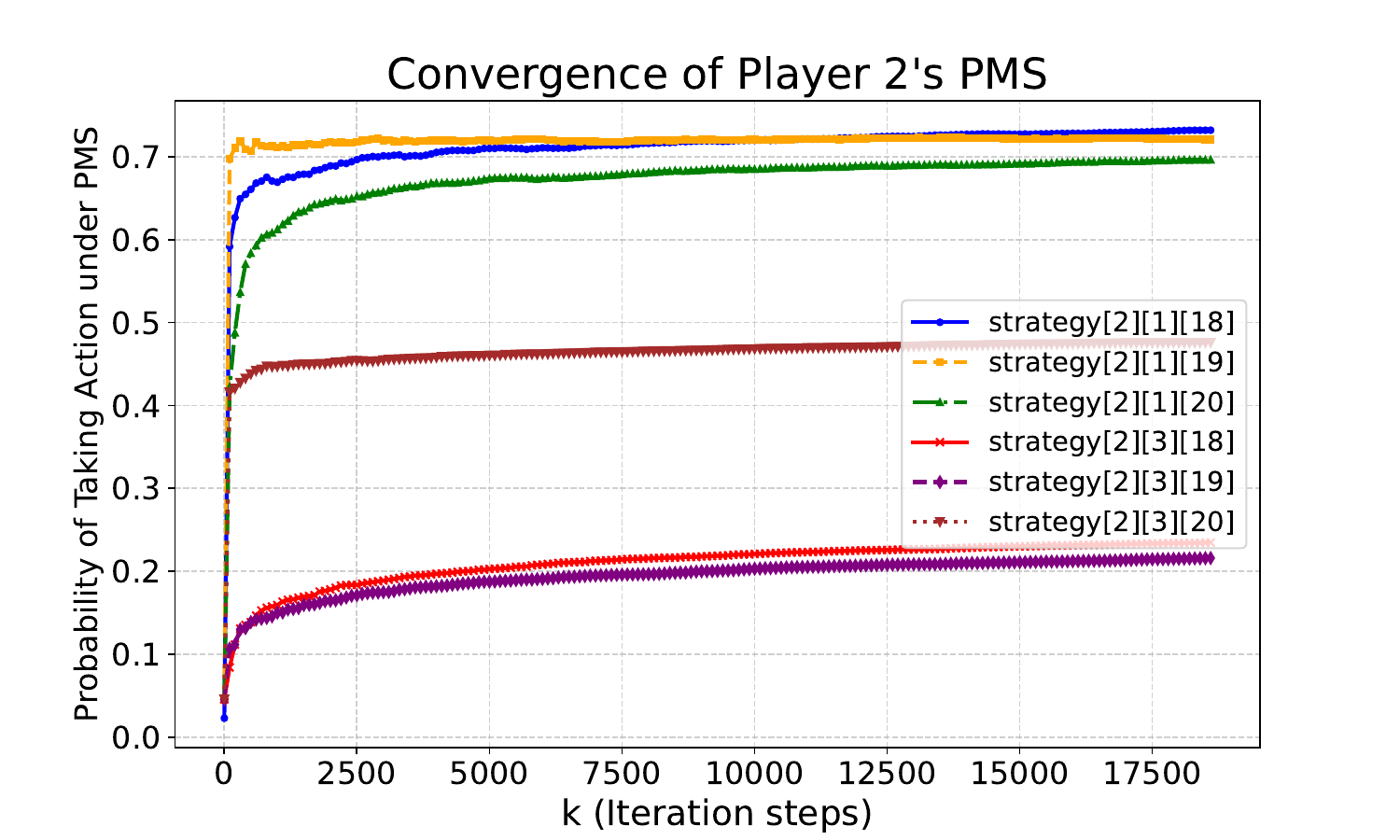}
	\end{subfigure}

	\vspace{0.05cm}  

	\begin{subfigure}{0.492\linewidth}
		\centering
		\includegraphics[width=\linewidth,trim=1.2cm 0.1cm 2.2cm 0.8cm, clip]{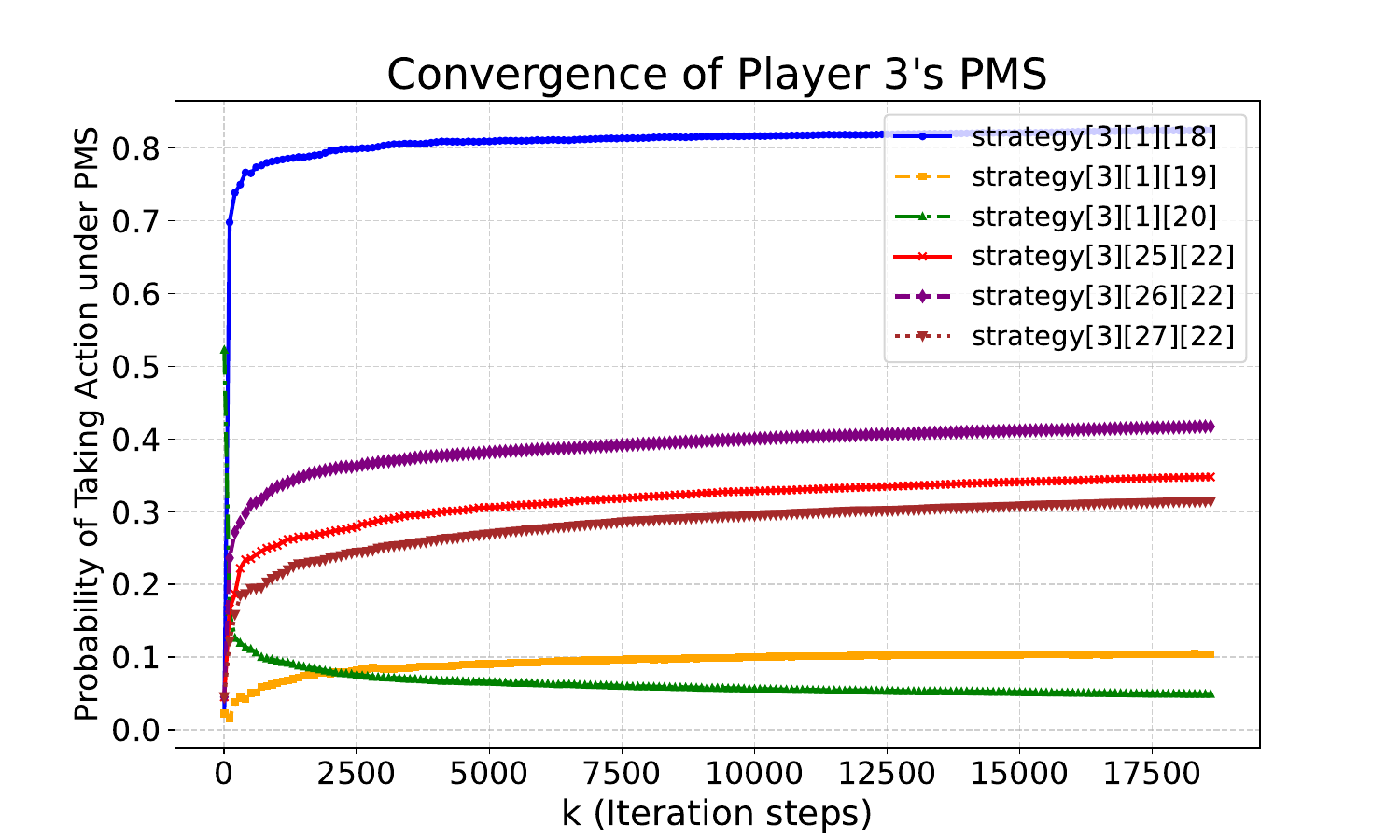}
	\end{subfigure}
	\hfill
	\begin{subfigure}{0.492\linewidth}
		\centering
		\includegraphics[width=\linewidth,trim=1.2cm 0.1cm 2.2cm 0.8cm, clip]{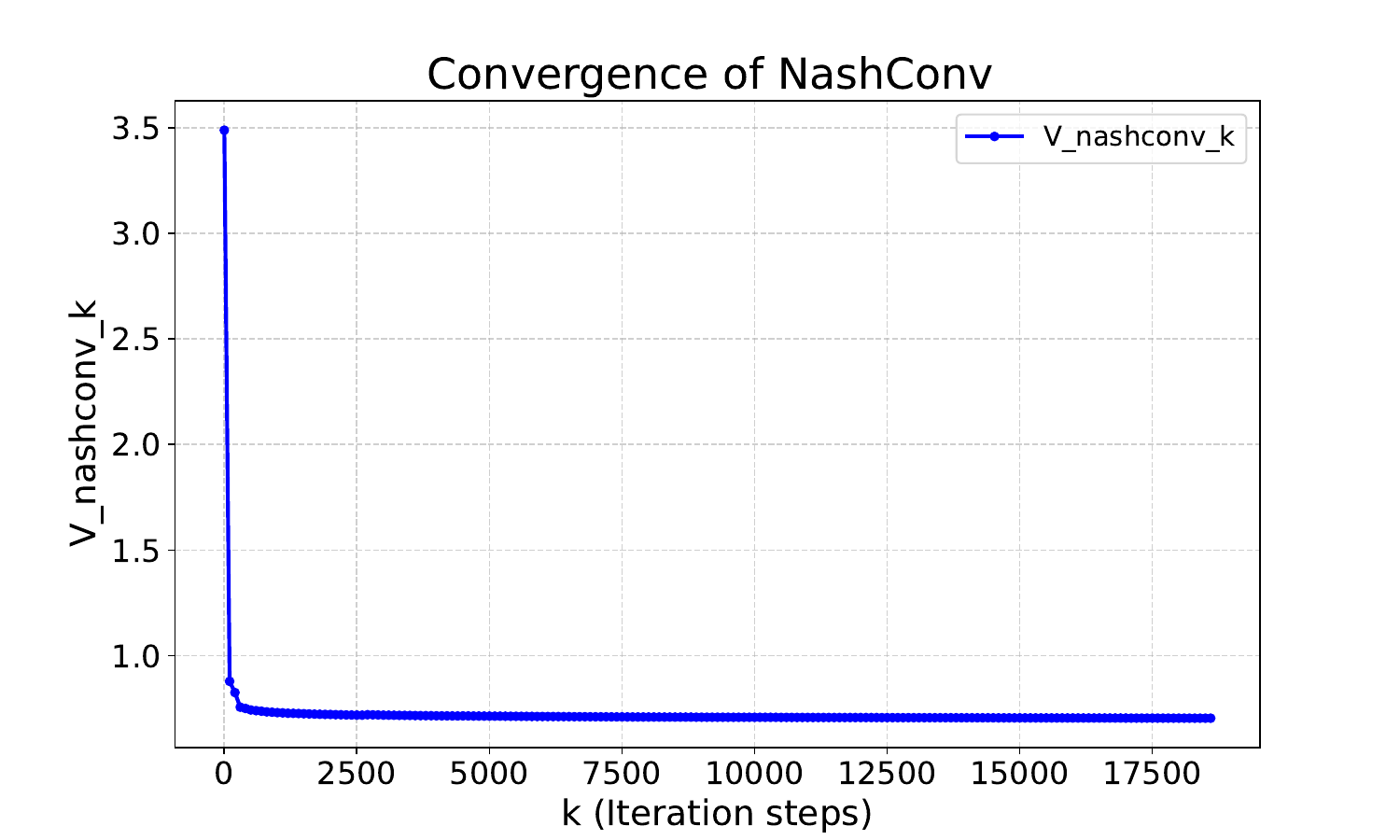}
	\end{subfigure}

	\caption{Iterative strategy adjustment and NashConv decay in FP+MDP.}
	\label{FP+MDP picture}
\end{figure}

\end{example}

To sum up, we propose two algorithms for computing equilibrium in the lower-level MG $\mathcal{G}_1$, each suitable for different scenarios:
\begin{enumerate}
    \item The \textit{FIP-based} algorithm is a centralized method designed for the constructed MPG. It requires all users' strategies and enables efficient and scalable equilibrium computation in large-scale, multi-user, long-horizon games.
    \item The \textit{FP+MDP} algorithm is a decentralized learning procedure applicable to general MGs. Users estimate others' strategies based solely on public information, making it suitable for MGs under incomplete information environments.
\end{enumerate}

\subsection{Solving the Optimal Pricing for the Upper-Level Aggregator}\label{4.2}
Having computed the equilibrium of $\mathcal{G}_1$, we now turn to the upper-level pricing optimization problem faced by the aggregator in the SMG.

For tractability in analysis and computation, we fix $\gamma_{1}$ and $\gamma_{2}$, and focus on optimizing the pricing parameters $(\alpha,\beta)$. Formally, the aggregator seeks $(\alpha^*, \beta^*)$ that maximize its expected payoff $U(\alpha, \beta, \boldsymbol{\mathbbm{1}}_{\boldsymbol{d}^*})$, where $\boldsymbol{\mathbbm{1}}_{\boldsymbol{d}^*}$ is the users' equilibrium under the parameters $(\alpha, \beta)$. Since the action space for $(\alpha, \beta)$ is discrete and finite, this problem can be solved through \textit{search}. The complete procedure is summarized in Algorithm~\ref{UL-optimization}.
\begin{algorithm}[H]
\caption{Upper-Level Pricing Optimization.}
\label{UL-optimization}
\begin{algorithmic}
\STATE \textbf{Input:} Discrete pricing action sets $\alpha \in \{\underline{\alpha}, \dots, \overline{\alpha}\}$, \STATE \hspace{1.0cm} $\beta \in \{\underline{\beta}, \dots, \overline{\beta}\}$.
\STATE \textbf{Initialization:}
\STATE \hspace{0.5cm} Set optimal value $U^* \gets -\infty$.
\STATE \hspace{0.5cm} Set optimal pricing $(\alpha^*, \beta^*) \gets (\underline{\alpha}, \underline{\beta})$.
\STATE \textbf{For each} $(\alpha, \beta)$ in the joint pricing action space \textbf{do}
\STATE \hspace{0.5cm} Compute $\boldsymbol{\mathbbm{1}}_{\boldsymbol{d}^*}$ via Algorithm~\ref{FIP}.
\STATE \hspace{0.5cm} Compute leader’s payoff $U(\alpha, \beta, \boldsymbol{\mathbbm{1}}_{\boldsymbol{d}^*})$.
\STATE \hspace{0.5cm} \textbf{if} $U(\alpha, \beta, \boldsymbol{\mathbbm{1}}_{\boldsymbol{d}^*}) > U^*$ \textbf{then}
\STATE \hspace{1.0cm} Update $U^* \gets U(\alpha, \beta, \boldsymbol{\mathbbm{1}}_{\boldsymbol{d}^*})$.
\STATE \hspace{1.0cm} Update $(\alpha^*, \beta^*) \gets (\alpha, \beta)$.
\STATE \hspace{0.5cm} \textbf{end if}
\STATE \textbf{end for}
\STATE \textbf{Output:} Optimal pricing $(\alpha^*, \beta^*)$ and corresponding equilibrium $\boldsymbol{\mathbbm{1}}_{\boldsymbol{d}^*}$.
\end{algorithmic}
\end{algorithm}

We continue with the setting of Example~\ref{FIP example} to analyze the upper-level pricing decisions.
\begin{example}[Example~\ref{FIP example} continued]\label{SMG eg}
Consider the MG from Example~\ref{FIP example}, where 50 users interact over 7 stages. The leader selects pricing parameters $\alpha, \beta \in \{19, 20, 21\}$ to maximize its expected payoff $U$ defined in \eqref{aggregator_payoff}, with parameters $C = 1$, $k = 0.1$, and $r_0 = 70$. Then the leader applies Algorithm~\ref{UL-optimization} to solve the optimal $(\alpha^*, \beta^*)$.
\begin{figure}[htb]
    \centering
    \includegraphics[width=\columnwidth]{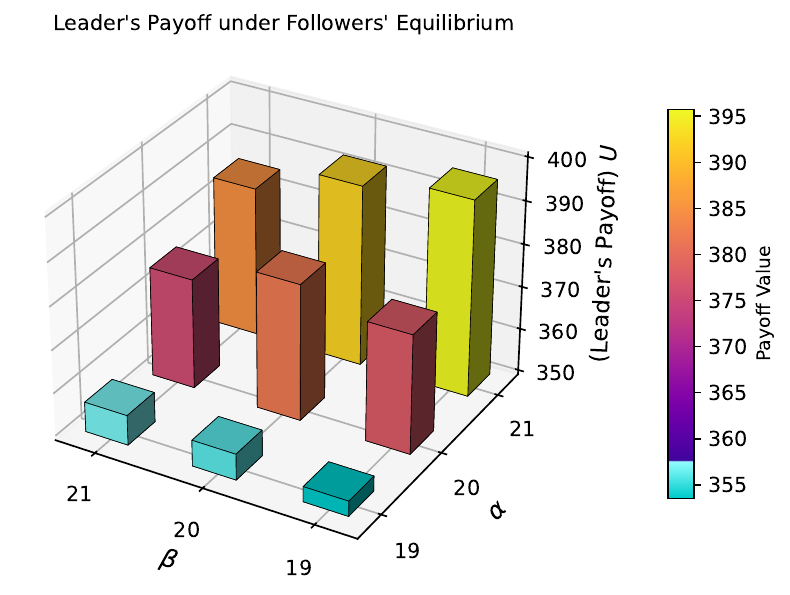}
    \caption{Optimization of Leader’s Payoff in the SMG.}
    \label{SMG example}
\end{figure}

As shown in Fig.~\ref{SMG example}, the leader's payoff varies across different pricing combinations.  
First, given $\beta$, increasing $\alpha$ raises the electricity price and reduces aggregate demand, but the overall profit still tends to increase.  
Second, the optimal choice of $\beta$ depends on $\alpha$: the leader's payoff is maximized at $\beta = 21$ when $\alpha = 19$, $\beta = 20$ when $\alpha = 20$, and $\beta = 19$ when $\alpha = 21$. 
The leader achieves the highest payoff at $(\alpha^*, \beta^*) = (21, 19)$.

Therefore, Algorithm~\ref{UL-optimization} effectively identifies the leader's optimal pricing strategy.
\end{example}

\subsection{Extension to Nonlinear Consumption Utility}\label{4.3}
In many real-world scenarios, users’ utility functions are nonlinear with respect to electricity consumption. 

To explore such cases, we replace the linear term in \eqref{current reward} with a piecewise-linear utility function $H_i(c_i^t)$ defined in \eqref{nonlinear}, which captures diminishing marginal benefits. Under this setting, we present an example where energy storage strategies strictly dominate non-storage strategies, illustrating richer user behaviors arising from more realistic utility modeling.

\begin{example}[Storage dominance under nonlinear utility]\label{Usefulness of Energy Storage} 
Consider the 2-user, 2-horizon MG defined by \eqref{G1}, where $\mathcal{T}=\mathcal{N} =\{1, 2\}$. The state space is $\{\mathcal{S}^t\}_{t=1}^{T=2}$ where $\mathcal{S}^t = \mathcal{E}_N^t \times \prod_{i=1}^{n} \mathcal{B}_{i}$ with $\mathcal E^1_{N} =\{ 10\}, \mathcal E^2_{N} =\{ 1,2,3\}$ and $\mathcal{B}_{i} =\{ 0,1\},\text{for}~ i=1,2$. Each user's action space is $\mathcal{A}_{i}$ with $\mathcal D_{i} = \{ 1,2,3\}$ and $\mathcal C_{i}=\{ 0,1,2,3,4\}$.  The reward function $r_i(\boldsymbol{s}^t,\boldsymbol{a}^t)$ (for $i=1,2$) follows \eqref{current reward}, except that the linear term $\theta_i  c_i^t$ is replaced by a utility function for electricity consumption, defined as: 
 \begin{align}\label{nonlinear}
H_{i}(c_{i}^{t}) = \begin{cases}
0.9  c_{i}^{t},  & \text{if } c_{i}^{t} \in \{0, 1, 2\},\\
1.8,  &\text{otherwise}.
\end{cases}
 \end{align}
Additionally, the parameters in the electricity pricing function are given by $\alpha = 1, \beta =1.5, \gamma_1=\gamma_2 = 1$.

The state transition probabilities $q(\cdot \mid \boldsymbol s^1,\boldsymbol a^1)$ are defined such that: 
\begin{align*}
q(\boldsymbol s^{2} &= (e^{2}_{N}, b^{2}_1, b^{2}_2)  \mid  \boldsymbol s^1 = (e^1_{N}  = 10, b^1_1=0,b^1_2=0),\boldsymbol a^t)  \notag \\
&= \begin{cases}
0.3,  & \text{if } e^{2}_{N}=1, b^{2}_{i}=d^{1}_{i}-c^{1}_{i}, \text{for } i=1,2,\\
0.4,  & \text{if } e^{2}_{N}=2, b^{2}_{i}=d^{1}_{i}-c^{1}_{i},  \text{for } i=1,2, \\
0.3,  & \text{if } e^{2}_{N}=3, b^{2}_{i}=d^{1}_{i}-c^{1}_{i},  \text{for } i=1,2.
\end{cases}
\end{align*}

For the MG described above, it can be verified that if user $i (i=1,2)$ adopts the energy storage strategy 
$\boldsymbol{\mathbbm{1}}_{\boldsymbol{ \bar a}_{i}} :=\{ \mathbbm{1}_{{ \bar a}^1_{i}} (e_{N}^1=10,b_i^1=0) =(\bar d_i^1=3,\bar c_i^1=2);\mathbbm{1}_{{ \bar a}^2_{i}} (e_{N}^2,b_i^2) =(\bar d_i^2=1,\bar c_i^2=2),\forall e_{N}^2 \in \mathcal E^2_{N}, b_i^2 \in \mathcal{B}_{i}\}$, this strategy strictly dominates any non-storage strategy $\boldsymbol{\mathbbm{1}}_{\hat{\boldsymbol{a}}_{i}} \in \mathcal{P}_{i}^{\boldsymbol d,T} := \{\boldsymbol{\mathbbm{1}}_{\boldsymbol{a}_{i}} \in \mathcal{P}^{PMS}_{i}\mid c^t_i = d^t_i \in \{1,2,3\} ,\forall t =1,2 \}$. The detailed verification process is provided in Appendix \ref{appendix3}.

\end{example}

\section{Conclusion}\label{conclusection}
In this paper, we proposed a novel Stackelberg Markov Game (SMG) framework to model the hierarchical interactions between an aggregator and multiple energy storage users under renewable generation uncertainty in the smart grid. The lower-level user interaction was formulated as a finite-horizon MG with Private Markovian Strategy (PMS). We provided a constructive proof of the existence of a Private Markovian Equilibrium (PME) in the specific lower-level game arising from our model. Furthermore, we proved that this PME can be computed in polynomial time by converting the PME problem into the ME problem of a constructed Markov potential game (MPG), thereby significantly simplifying equilibrium computation. Building on this, we developed scalable equilibrium solution algorithms under complete and incomplete information settings, and further designed a search-based method to determine the leader’s optimal pricing strategy based on the computed equilibrium. We validated the effectiveness and scalability of the proposed methods through extensive simulations. The results provide theoretical support for real-time pricing (RTP) mechanism design in renewable-integrated smart grids.

There are many interesting and significant future directions.
First, it remains an open question whether the SMG equilibrium can still be efficiently computed when more complex payoff structures are considered, and whether the conversion to MPG remains applicable.
Second, incorporating more realistic user behaviors, such as the ability to generate electricity, non-negligible storage costs, or imperfect charging/discharging efficiencies, would further enrich the model and lead to new insights into system dynamics.
Third, when solving for exact equilibrium in the user-level MG becomes intractable, exploring approximate or relaxed solution concepts as substitutes could be a promising direction for ensuring the tractability of the overall SMG framework.

\bibliographystyle{IEEEtran}
\bibliography{refs}

\appendix
\section{On the Existence of Private Markovian Equilibrium}\label{appendix_pme} 
We consider a class of MG where the system state at time $t$ is $\boldsymbol s^t = (s_{\mathrm{pub}}^t, b_1^t, \dots, b_n^t) \in \mathcal{S}^t$, with $s_{\mathrm{pub}}^t$ denoting the public state and $b_i^t$ representing player $i$’s individual state. In the PMS $\pi_i$, player $i$’s decision at time $t$ depends only on private state $\boldsymbol s^t_i :=(s_{\mathrm{pub}}^t, b_i^t) \in \mathcal{S}_i^t$, and not on the individual states of other players. This restriction reduces the admissible strategy space compared to general Markovian strategies, which may condition on the full joint state.

As a result, classical existence theorems for ME, which rely on strategies defined over the full state space, do not directly apply under PMS. Moreover, the relationship between PME and ME is not one of strict inclusion: although PMS imposes stronger locality constraints, the equilibrium sets of PME and ME are not necessarily nested. Therefore, PME should not be viewed as a refinement or subset of ME in the standard sense.

Despite these challenges, PME remains a meaningful equilibrium concept in decentralized settings, where agents typically have access only to public information and their own local states. PMS naturally captures this form of partial observability, and analyzing PME helps quantify the role of information structures in equilibrium behavior.

The following proposition establishes the existence of PME under certain sufficient conditions in the finite-horizon MG.

\begin{proposition}\label{SE in the G_2}
A private Markovian equilibrium (PME) exists in the finite-horizon Markov game if either of the following conditions holds:
\begin{enumerate}
    \item The MG is a game with perfect recall under the information structure induced by private Markovian strategies (PMS);
    \item The value function $V^{\boldsymbol \pi_i,\boldsymbol \pi_{-i}}_{1,i}(\boldsymbol s^1)$ is quasi-concave in $\boldsymbol \pi_i$ for each fixed $\boldsymbol \pi_{-i}$ and all initial states $\boldsymbol s^{1} \in \boldsymbol S^1, i \in \mathcal{N}$.
\end{enumerate}
\end{proposition}
{\it Proof:} 
We will prove each condition in turn.

\textit{1.} Fix any initial state $\boldsymbol s^1 \in \mathcal{S}^1$. Under the PMS strategy structure, the finite-horizon game can be represented as a finite extensive-form game where information sets for each player $i$ are determined by their observed local state $(s_{\mathrm{pub}}^t, b_i^t)$. By assumption, these information sets satisfy perfect recall.

Because the number of players, states, actions, and time steps is finite, the total number of decision nodes and hence the number of information sets in the extensive-form game is also finite. Consequently, each player has a finite set of pure PMS. Thus, the game can be equivalently represented as a strategic-form game with finite players and finite pure strategies. By Nash’s theorem, an equilibrium in mixed strategies exists. Furthermore, by Kuhn’s theorem, since the game has perfect recall, there exists an equilibrium in behavioral strategies, which correspond to PMS strategies. Therefore, there exists a private Markovian equilibrium (PME) in the $T$-horizon MG.

\textit{2.} Consider each PMS strategy $\boldsymbol \pi_i: \cup_{t=1}^{T}\{ t \} \times \mathcal{S}_i^t \rightarrow \cup_{t=1}^{T}\Delta(\mathcal{A}_i^t)$. The PMS strategy space $\Pi_i$ is a finite Cartesian product of simplices, hence compact and convex.

Since $V^{\boldsymbol \pi_i,\boldsymbol \pi_{-i}}_{1,i}(\boldsymbol s^1)$ is a multivariate polynomial in the PMS profile $\boldsymbol \pi$, it is continuous. By assumption, $V^{\boldsymbol \pi_i,\boldsymbol \pi_{-i}}_{1,i}(\boldsymbol s^1)$ is quasi-concave in $\boldsymbol \pi_i$ for each fixed $\boldsymbol \pi_{-i}$, so the best response correspondence $BR_i(\boldsymbol \pi_{-i}) := \arg\max_{\boldsymbol \pi_i \in \Pi_i} V^{\boldsymbol \pi_i,\boldsymbol \pi_{-i}}_{1,i}(s^1)$ is non-empty, convex-valued, and upper hemicontinuous.

The joint best response correspondence $\boldsymbol{BR}(\boldsymbol \pi) := \times_{i\in \mathcal{N}} BR_i(\boldsymbol \pi_{-i})$, as a finite product of such correspondences, also inherits these properties. Since the joint strategy space $\Pi = \times_{i\in \mathcal{N}} \Pi_i$ is compact and convex, Kakutani’s fixed point theorem guarantees the existence of a fixed point $\boldsymbol \pi^* \in \boldsymbol{BR}(\boldsymbol \pi^*)$, which constitutes a PME.
\te

While these conditions provide insight into structured cases where PME exists, a general characterization of PME existence remains an open problem. Future research may explore relaxed strategy spaces or approximation frameworks to establish broader existence guarantees under partial observability.

The following gives the formal definition of a game with perfect recall.

\begin{definition}[Perfect recall] A game is called a game with perfect recall if all the players have perfect recall. Player $i$ is said to have perfect recall if the following conditions are satisfied:
\begin{enumerate}
	\item \label{condition1} Every information set of player $i$ intersects each path from the root to a leaf at most once.   
       \item \label{condition2}Every two paths from the root that end in the same information set of player $i$ pass through the same information sets of player $i$, and in the same order, and in every such information set the two paths choose the same action. In other words, for every
information set $U_i$ of player $i$ and every pair of vertices $x, \hat{x} \in U_i$, if the decision
vertices of player $i$ on the path from the root to $x$ and $\hat{x}$ are $x^1_i,x^2_i, \cdots, x^L_i = x$ and $\hat x^1_i,\hat x^2_i, \cdots, \hat x^{\hat L}_i = \hat x$,  respectively, then $L =\hat L$, and $U_{i}\left(x_{i}^{l}\right)=U_{i}\left(\widehat{x}_{i}^{l}\right)$, and  $a_{i}\left(x_{i}^{l} \rightarrow x\right)=a_{i}\left(\widehat{x}_{i}^{l} \rightarrow \widehat{x}\right)$ for all $l \in\{1,2, \ldots, L\}$ .
\end{enumerate}
These conditions ensure that each player remembers their own past decisions and observations, forming the basis of the game's perfect recall property.
\end{definition}

\section{The Equivalence of the Definitions}\label{appendix2}
\begin{lemma}\label{equivalent of potential definition}
The following equivalences hold:
\begin{enumerate}
    \item The notions of potential games in pure strategies and in mixed strategies, as given in Definition~\ref{Potential Game}, are equivalent;
    \item The notions of MPGs in pure Markovian strategies and in general Markovian strategies, as given in Definition~\ref{Stochastic Potential Game 1}, are equivalent.
\end{enumerate}
\end{lemma}
{\it Proof:} 
We prove each equivalence in turn.

\textit{1.}  Clearly, \eqref{potential game function 1} implies \eqref{potential game function}. Conversely, if \eqref{potential game function} holds, then
    \begin{align}\label{equivalent of potential game function}
    &u_{i}\left(m_{i}, m_{-i}\right) - u_{i}\left(m_{i}^{\prime}, m_{-i}\right) \notag \\
    &= \sum_{x_i \in \mathcal{X}_i} \left(m_i(x_i) - m_{i}^{\prime}(x_i)\right) 
    \Bigg[\sum_{x_{-i} \in \mathcal{X}_{-i}} m_{-i}(x_{-i}) u_i(x_i,x_{-i}) \Bigg]  \notag \\
    & = \sum_{x_i \in \mathcal{X}_i\setminus \bar x_i} \left(m_i(x_i) - m_{i}^{\prime}(x_i)\right) 
    \Bigg[\sum_{x_{-i} \in \mathcal{X}_{-i}} m_{-i}(x_{-i})\Big[u_i(x_i,x_{-i}) - u_i(\bar x_i,x_{-i}) \Big] \Bigg] 
    \notag \\
    &= \sum_{x_i \in \mathcal{X}_i \setminus \bar x_i} \left(m_i(x_i) - m_{i}^{\prime}(x_i)\right) 
    \Bigg[\sum_{x_{-i} \in \mathcal{X}_{-i}} m_{-i}(x_{-i})\Big[\phi(x_i,x_{-i}) - \phi(\bar x_i,x_{-i}) \Big] \Bigg]  \notag \\
    & =  \sum_{x_i \in \mathcal{X}_i} \left(m_i(x_i) - m_{i}^{\prime}(x_i)\right) 
    \Bigg[\sum_{x_{-i} \in \mathcal{X}_{-i}} m_{-i}(x_{-i})\phi(x_i,x_{-i}) \Bigg] \notag \\
    & = \phi\left(m_{i}, m_{-i}\right) - \phi\left(m_{i}^{\prime}, m_{-i}\right),~\forall i \in \mathcal{N},m_{i}, m_{i}^{\prime} \in \mathcal{M}_{i}, m_{-i} \in \mathcal{M}_{-i}.
    \end{align}
    Therefore, \eqref{potential game function} implies \eqref{potential game function 1}.
    
\textit{2.}  Clearly, \eqref{Stochastic potential game function} implies \eqref{Stochastic potential game function 1} because a pure strategy profile $\boldsymbol{\mathbbm{1}}_{\boldsymbol{a}}$ can be regarded as a degenerate Markovian strategy profile. 

    Conversely, suppose that \eqref{Stochastic potential game function 1} holds.
    Since a general finite-horizon MG is a game with perfect recall (see Remark 15.8 in \cite{maschler2020game}), according to Kuhn's Theorem \cite{kuhn1953extensive}, every mixed strategy has a strategically equivalent behavioral strategy (and vice versa). Therefore, consider $(\sigma_i,\sigma_{-i})$ and $(\sigma_i^{\prime},\sigma_{-i})$ are two behavior strategies or Markovian strategies, and let $(\tau_i,\tau_{-i})$ and $(\tau_i^{\prime},\tau_{-i})$ be the strategically equivalent mixed strategies, respectively. Here, we denote the mixed strategies $\tau_i := \Delta(\{\boldsymbol{\mathbbm{1}}_{\boldsymbol{a_i}}:\boldsymbol{a_i} \in \mathcal{A}_i \})$, i.e., probability distributions over pure Markovian strategies.

    Then, the expected payoff satisfies:
    \begin{align*}
        V^{\sigma_i,\sigma_{-i}}_{1,i}(\boldsymbol s^{1}) -V^{\sigma_i^{\prime},\sigma_{-i}}_{1,i}(\boldsymbol s^{1}) = V^{\tau_i,\tau_{-i}}_{1,i}(\boldsymbol s^{1}) -V^{\tau_i^{\prime},\tau_{-i}}_{1,i}(\boldsymbol s^{1}), ~\forall i \in \mathcal{N}, \boldsymbol s^1 \in \boldsymbol S^1.
    \end{align*}
    
    Now we analyze the change in player $i$’s payoff under mixed strategies. Based on \eqref{equivalent of potential game function}, we can derive a similar conclusion:  
\begin{align*}
    &V^{\tau_i,\tau_{-i}}_{1,i}(\boldsymbol s^{1}) -V^{\tau_i^{\prime},\tau_{-i}}_{1,i}(\boldsymbol s^{1})  \\
    &= \sum_{\boldsymbol{a_i} \in \mathcal{A}_i} \left(\tau_i(\boldsymbol{\mathbbm{1}}_{\boldsymbol{a_i}}) - \tau_{i}^{\prime}(\boldsymbol{\mathbbm{1}}_{\boldsymbol{a_i}})\right) 
    \Bigg[\sum_{\boldsymbol a_{-i} \in \mathcal{A}_{-i}} \tau_{-i}(\boldsymbol{\mathbbm{1}}_{\boldsymbol{a}_{-i}})  V^{\boldsymbol{\mathbbm{1}}_{\boldsymbol{a}_{i}},\boldsymbol{\mathbbm{1}}_{\boldsymbol{a}_{-i}}}_{1,i}(\boldsymbol s^{1}) \Bigg]   \\
    & = \sum_{\boldsymbol{a_i} \in \mathcal{A}_i\setminus \boldsymbol{\bar a_i}} \left(\tau_i(\boldsymbol{\mathbbm{1}}_{\boldsymbol{a_i}}) - \tau_{i}^{\prime}(\boldsymbol{\mathbbm{1}}_{\boldsymbol{a_i}})\right) 
    \Bigg[\sum_{\boldsymbol a_{-i} \in \mathcal{A}_{-i}} \tau_{-i}(\boldsymbol{\mathbbm{1}}_{\boldsymbol{a}_{-i}}) \Big[V^{\boldsymbol{\mathbbm{1}}_{\boldsymbol{a}_{i}},\boldsymbol{\mathbbm{1}}_{\boldsymbol{a}_{-i}}}_{1,i}(\boldsymbol s^{1}) - V^{\boldsymbol{\mathbbm{1}}_{\boldsymbol{\bar a}_{i}},\boldsymbol{\mathbbm{1}}_{\boldsymbol{a}_{-i}}}_{1,i}(\boldsymbol s^{1})\Big] \Bigg] \\ 
    &= \sum_{\boldsymbol{a_i} \in \mathcal{A}_i\setminus \boldsymbol{\bar a_i}} \left(\tau_i(\boldsymbol{\mathbbm{1}}_{\boldsymbol{a_i}}) - \tau_{i}^{\prime}(\boldsymbol{\mathbbm{1}}_{\boldsymbol{a_i}})\right) 
    \Bigg[\sum_{\boldsymbol a_{-i} \in \mathcal{A}_{-i}} \tau_{-i}(\boldsymbol{\mathbbm{1}}_{\boldsymbol{a}_{-i}}) \Big[\phi_{\boldsymbol s^{1}}\left(\boldsymbol{\mathbbm{1}}_{\boldsymbol{a}_{i}},\boldsymbol{\mathbbm{1}}_{\boldsymbol{a}_{-i}}\right)-\phi_{\boldsymbol s^{1}} \left(\boldsymbol{\mathbbm{1}}_{\boldsymbol{\bar a}_i^{\prime}},\boldsymbol{\mathbbm{1}}_{\boldsymbol{a}_{-i}}\right)\Big] \Bigg]   \\
    & = \phi_{\boldsymbol s^{1}}\left(\tau_i,\tau_{-i}\right)-\phi_{\boldsymbol s^{1}} \left(\tau_i^{\prime},\tau_{-i}\right), ~\forall i \in \mathcal{N}, \boldsymbol s^1 \in \boldsymbol S^1, \tau_{i}, \tau_{i}^{\prime} \in \mathcal{T}_{i}, \tau_{-i} \in \mathcal{T}_{-i}.
    \end{align*}

    Finally, define the potential function value under behavior (Markov) strategies to be consistent with their mixed-strategy equivalents:
\begin{align*}
&\phi_{\boldsymbol s^{1}}\left(\sigma_i,\sigma_{-i}\right) = \phi_{\boldsymbol s^{1}}\left(\tau_i,\tau_{-i}\right), \phi_{\boldsymbol s^{1}}\left(\sigma_i^{\prime},\sigma_{-i}\right) = \phi_{\boldsymbol s^{1}}\left(\tau_i^{\prime},\tau_{-i}\right), ~\forall \boldsymbol s^1 \in \boldsymbol S^1.
\end{align*}

Consequently, it follows that:
\begin{align*}
V^{\sigma_i,\sigma_{-i}}_{1,i}(\boldsymbol s^{1}) -V^{\sigma_i^{\prime},\sigma_{-i}}_{1,i}(\boldsymbol s^{1})  = \phi_{\boldsymbol s^{1}}\left(\sigma_i,\sigma_{-i}\right)-\phi_{\boldsymbol s^{1}} \left(\sigma_i^{\prime},\sigma_{-i}\right), \notag \\\forall i \in \mathcal{N}, \boldsymbol s^1 \in \boldsymbol S^1,\sigma_i, \sigma_i^{\prime} \in \Sigma_{i}, \sigma_{-i} \in \Sigma_{-i}.
\end{align*}
Thus, \eqref{Stochastic potential game function 1} implies \eqref{Stochastic potential game function}.

\te

\section{Extension of MPGs under PMS}\label{appendixX}

The definitions of Markov Potential Games (MPGs) in Definition~\ref{Stochastic Potential Game 1} are based on Markovian strategies, where decisions depend solely on public state information. We now extend the concept of MPGs to the Private Markovian Strategy (PMS) framework, where each player's decision is based only on their private observation—comprising the public state and their own private state.
\begin{definition}\label{Stochastic potential game 3}
(Markov Potential Game under PMS). A finite-horizon Markov Game 
\[
\left\{\mathcal{T}, \mathcal{N}, \{\mathcal{S}^t\}_{t=1}^T, \left\{\mathcal{A}_{i}\right\}_{i=1}^{n}, \left\{{r}_{i}\right\}_{i=1}^{n}, q \right\}
\]  is a Markov Potential Game (MPG) under PMS if there exists a state-dependent potential function $\phi_{\boldsymbol{s}}: \mathcal{P} \rightarrow \mathbb{R}$ such that
\begin{align*} &V^{\boldsymbol{\mathbbm{1}}_{\boldsymbol{a}_{i}},\boldsymbol{\mathbbm{1}}_{\boldsymbol{a}_{-i}}}_{1,i}(\boldsymbol s^{1}) -V^{\boldsymbol{\mathbbm{1}}_{\boldsymbol{a}_i^{\prime}},\boldsymbol{\mathbbm{1}}_{\boldsymbol{a}_{-i}}}_{1,i}(\boldsymbol s^{1})= \phi_{\boldsymbol s^{1}}\left(\boldsymbol{\mathbbm{1}}_{\boldsymbol{a}_{i}},\boldsymbol{\mathbbm{1}}_{\boldsymbol{a}_{-i}}\right) -\phi_{\boldsymbol s^{1}} \left(\boldsymbol{\mathbbm{1}}_{\boldsymbol{a}_i^{\prime}},\boldsymbol{\mathbbm{1}}_{\boldsymbol{a}_{-i}}\right)
\end{align*}
for every  $i \in \mathcal{N}, \boldsymbol s^{1} \in \boldsymbol S^1,\boldsymbol{\mathbbm{1}}_{\boldsymbol{a}_{i}}, \boldsymbol{\mathbbm{1}}_{\boldsymbol{a}_i^{\prime}} \in \mathcal{P}_{i}$ , and  $ \boldsymbol{\mathbbm{1}}_{\boldsymbol{a}_{-i}} \in \mathcal{P}_{-i}$ .
\end{definition}
Here, the pure strategy $(\boldsymbol{\mathbbm{1}}_{\boldsymbol{a}_{i}},\boldsymbol{\mathbbm{1}}_{\boldsymbol{a}_{-i}})$ refers to a degenerate case of PMS, consistent with the notation in Theorem~\ref{main thm}.

If the PMS framework satisfied the perfect recall property, classical results from potential game theory—such as the equivalence between pure and mixed strategies and the existence of a pure-strategy equilibrium via potential maximization—would apply. In the absence of perfect recall, however, these results cannot be directly extended. Nonetheless, Definition~\ref{Stochastic potential game 3} still provides a meaningful structural foundation for analyzing games under PMS, although equilibrium properties must be studied separately.

\section{Proof of Lemma \ref{one-horizon}}\label{prop1}
{\it Proof:} We prove each item in turn.

\textit{1.} Without loss of generality, consider a single-stage MG at state $\boldsymbol s^{t} \in \mathcal S^t$. The game reduces to a strategic-form game 
$\mathcal{M}(\boldsymbol s^{t})=\langle \mathcal{N}, (\mathcal{A}_{i}^t)_{i \in \mathcal{N}}, (r_{i})_{i \in \mathcal{N}} \rangle$. 
If users take action $\boldsymbol{a}^t=(a^t_{i},a^t_{-i})\in \mathcal{A}^t$, then by \eqref{current reward}, player $i$'s payoff is:
\begin{multline}\label{linear reward}
r_i(\boldsymbol{s}^t,\boldsymbol{a}^t) =  \theta_i  c_{i}^t - P(d^{t}, e^t_{N})  d^{t}_{i}
= \theta_i  c_{i}^t - (\frac{\alpha}{n e^t_{N}+\gamma_{1}} d^{t}+\frac{\beta} {e^t_{N}+\gamma_{2}})  d^{t}_{i},~ \forall \boldsymbol{s}^t \in \mathcal{S}^t.
\end{multline}

Considering the following function:
\begin{multline*}
     \bar \phi_{\boldsymbol{s}^t}(\boldsymbol{a}^t)  = \sum_{i=1}^{n} \theta_i  c_{i}^t- \frac{\beta} {e^t_{N}+\gamma_{2}} d^t -\frac{\alpha}{n e^t_{N}+\gamma_{1}} 
      \sum_{i=1}^{n} (d_{i}^{t})^2 -\frac{\alpha}{n e^t_{N}+\gamma_{1}} \sum_{1 \leq i<j \leq n} d_{i}^{t}  d_{j}^{t} ,~ \forall \boldsymbol{s}^t \in \mathcal{S}^t,
\end{multline*}
It is easy to check that $\bar \phi_{\boldsymbol{s}^t}(\cdot)$ satisfies the exact potential game condition \eqref{potential game function}, since:
\begin{align*}
    &r_i(\boldsymbol{s}^t,a^t_{i},a^t_{-i}) -r_i(\boldsymbol{s}^t,a^{\prime t}_{i},a^t_{-i})  \\
    &=\theta_i  (c_{i}^t- c_{i}^{\prime t})-\frac{\beta} {e^t_{N}+\gamma_{2}} (d_{i}^t- d_{i}^{\prime t}) - \frac{\alpha}{n e^t_{N}+\gamma_{1}} \left[(d_{i}^t)^2-(d_{i}^{\prime t})^2 + (d_{i}^t-d_{i}^{\prime t}) d_{-i}^{t}\right], \\
     &=\bar \phi_{\boldsymbol{s}^t}(a^t_{i},a^t_{-i}) - \bar \phi_{\boldsymbol{s}^t}(a^{\prime t}_{i},a^t_{-i}),
\end{align*}
for all $i \in \mathcal{N}$, $\boldsymbol{s}^t \in \mathcal{S}^t$, $a^t_{i}, a^{\prime t}_{i} \in \mathcal{A}_{i}^t$, and $a^t_{-i} \in \mathcal{A}_{-i}^t$.

Thus, the single-stage MG is an exact potential game.

\textit{2.}
Consider again the strategic-form game $\mathcal{M}(\boldsymbol{s}^{t})$ as in item 1), where each user's action space depends on $\boldsymbol{s}^{t}$ due to \eqref{relationship}, i.e., $\mathcal{A}^t_i(\boldsymbol{s}^t) = \{{a}^{t}_{i}=( {d}^{t}_{i},{c}^{t}_{i}) \mid d_{i}^t \in \mathcal D_{i}, c_{i}^t \in \mathcal C_{i} \cap [b^t_{i} +  d^t_{i} - b^{\text{max}}_{i},  b^t_{i} +  d^t_{i}]\}$. Here, the feasible set of $d_{i}^t$ is state-independent, while $c_{i}^t$ is constrained by $b_{i}^t$, a component of state $\boldsymbol{s}^t$.

Since electricity consumption $c^t_{i}$ appears linearly in the payoff function \eqref{linear reward}, any action with $c^t_{i}<d^t_{i} + b^t_i$ is strictly dominated. We define a reduced action set excluding dominated actions:
\begin{align}\label{single-A}
\mathcal{\bar A}^t_i(\boldsymbol s^{t}) := \{\bar a^t_i=(d^t_{i},c^t_{i}) \mid {a}^t_i \in \mathcal{A}^t_i(\boldsymbol s^{t}) , c^t_{i}=d^t_{i} + b^t_i\},
\end{align}
Under this restriction, user $i$'s payoff becomes:
\begin{align}\label{after_excluding_dominated_actions}
r_i(\boldsymbol{s}^t,\boldsymbol{\bar a}_i^t, \boldsymbol{a}_{-i}^t) = \theta_i  (d^t_{i} + b^t_i) - (\frac{\alpha}{n e^t_{N} +\gamma_{1}} d^{t}  +\frac{\beta} {e^t_{N}+\gamma_{2}})  d^{t}_{i}.
\end{align}

For a fixed state $\boldsymbol{s}^{t} = (e^t_{N}, b^t_1, \dots, b^t_n)$ and given actions $\boldsymbol{a}^t_{-i}$ of the other users, the term $\theta_i b^t_i$ in \eqref{after_excluding_dominated_actions} is constant with respect to user~$i$'s decision after excluding dominated actions. Hence, the payoff depends only on $d^t_i$, and the best-response problem can be written as:
\begin{align}\label{max_g}
&\max \limits_{\boldsymbol{a}^t_{i} \in \mathcal{A}^t_i(\boldsymbol s^{t})} r_i(\boldsymbol{s}^t,\boldsymbol{a}^t_i,\boldsymbol{a}^t_{i}) = \max \limits_{\boldsymbol{\bar a}_{i}^t \in \mathcal{\bar A}^t_i(\boldsymbol s^{t})} r_i(\boldsymbol{s}^t,\boldsymbol{\bar a}_i^t, \boldsymbol{a}_{-i}^t)  = \max \limits_{d_{i}^t \in \mathcal D_{i}} g_i(e^t_{N},\boldsymbol{d}^t),
\end{align}
where $ g_i(e^t_{N},\boldsymbol{d}^t) := \theta_i  d^t_{i} - P(d^{t}, e^t_{N})  d^{t}_{i}$ denotes the stage payoff that depends solely on the electricity demand with $\boldsymbol{d}^t := (d_i^t)_{i \in \mathcal{N}}$.

This leads to a reduced game in which each user chooses \( d^t_i \in \mathcal{D}_i \), with payoff given by $g_i(e^t_N, \boldsymbol{d}^t)$. It is easy to verify that this reduced game admits an exact potential function:
\begin{multline*}
\phi(e^t_{N}, \boldsymbol{d}^{t}) =  \sum_{i=1}^{n} (\theta_i-\frac{\beta} {e^t_{N}+\gamma_{2}}) d_{i}^t-\frac{\alpha}{n e^t_{N}+\gamma_{1}}  
\sum_{i=1}^{n} (d_{i}^t)^2 -\frac{\alpha}{n e^t_{N}+\gamma_{1}} \sum_{1 \leq i<j \leq n} d_{i}^t  d_{j}^t, ~ \forall e^t_{N} \in \mathcal E^t_{N}.
\end{multline*}
By \cite{monderer1996potential}, any finite potential game has a pure Nash equilibrium, denoted by
$\boldsymbol{d}^{*t}(e^t_{N}) = \arg \max_{\boldsymbol{d}^{t} \in \mathcal D} \phi_{e^t_{N}}(\boldsymbol{d}^t)$. If not unique, we fix an arbitrary selection.

Since the equilibrium strategies in the reduced game determine the optimal choices of $d_i^t$, and $c_i^t$ is uniquely determined by $d_i^t$ and $b_i^t$ after excluding strictly dominated actions, we obtain a PME for the original game $\mathcal{M}(\boldsymbol{s}^{t})$ as:
\begin{align*}
\boldsymbol{a}^{*t}_i(\boldsymbol s_i^{t}) = \left(d_i^{*t}(e^t_{N}), d_i^{*t}(e^t_{N})+ b^t_i \right), ~ \forall i \in \mathcal{N}.
\end{align*}

Therefore, the equilibrium demand $d_i^{*t}(e^t_{N})$ depends only on $e^t_{N}$. Specifically, for any two states \( \boldsymbol{s}^{t} \) and \( \boldsymbol{s}^{\prime t} \) such that \( e^t_{N} = e^{\prime t}_{N} \), the pure equilibrium demands in these games are identical and given by $ \boldsymbol{d}^{*t}(e^t_{N})$. This completes the proof. 
\te

\section{Proof of Theorem \ref{main thm}}\label{thm_1}
{\it Proof:}
To solve the PME, we analyze the best response of user~$i$ given the pure PMS profile of other users, starting from any initial state $ \boldsymbol s^1 \in \mathcal S^1$.

Recall that a pure PMS of user $i$ is a mapping 
$\boldsymbol{\mathbbm{1}}_{\boldsymbol{a}_i}(\cdot) \in \mathcal{P}^{PMS}_i$ such that $\mathbbm{1}_{{a}^t_i}(\boldsymbol{s}_i^t) \in \mathcal{A}_i^t$ for all $t \in \mathcal{T}, \boldsymbol{s}_i^t \in \mathcal{S}^t_i$.

Given the pure PMS $\boldsymbol{\mathbbm{1}}_{\boldsymbol{a}_{-i}}$ of other users, user $i$ can use \textit{backward induction} to obtain their best response strategy. The procedure consists of the following four steps:

\textit{Step 1. Equilibrium Solving at Stage $T$:} Given any PMS $(\boldsymbol{\mathbbm{1}}_{\boldsymbol{a}_{i}},\boldsymbol{\mathbbm{1}}_{\boldsymbol{a}_{-i}})$, expand the expected value function \eqref{V_PMS} of user $i$ as follows:
 \begin{align}\label{Expanding the expected value-function}
    & V^{\boldsymbol{\mathbbm{1}}_{\boldsymbol{a}_{i}},\boldsymbol{\mathbbm{1}}_{\boldsymbol{a}_{-i}}}_{1,i}(\boldsymbol s^1) \notag \\   & =r_{i}\left(\boldsymbol s^{1}, \boldsymbol a^{1}\right) + \sum_{\boldsymbol s^{2} \in \mathcal{S}^2}q(\boldsymbol s^{2} \mid  \boldsymbol s^1, \boldsymbol a^{1})\Big[ r_{i}\left(\boldsymbol s^{2}, \boldsymbol a^{2}\right)   + \cdots  +   \sum_{\boldsymbol s^{T-1} \in \mathcal{S}^{T-1}}q(\boldsymbol s^{T-1} \mid  \boldsymbol s^{T-2}, \boldsymbol a^{T-2})\times  \notag \\   &\quad\quad \big[r_{i}\left(\boldsymbol s^{T-1}, \boldsymbol a^{T-1}\right) + \sum_{\boldsymbol s^{T} \in \mathcal{S}^{T}}q(\boldsymbol s^{T} \mid  \boldsymbol s^{T-1}, \boldsymbol a^{T-1}) [r_i(\boldsymbol{s}^T,\boldsymbol{a}^T)]\big]\cdots\Big]. 
\end{align} 

In \eqref{Expanding the expected value-function}, focus on $r_i(\boldsymbol{s}^T,\boldsymbol{a}^T)$, the stage payoff at stage $T$. Since $r_i(\boldsymbol{s}^T,\boldsymbol{a}^T) = \theta_i c_i^T - P(d^{T}, e^T_{N}) d_i^T$ is linear in $c^t_{i}$, user $i$ will always choose $c^T_{i} = d^T_{i}+ b^T_{i}$ to maximize his payoff, regardless of the actions of other users. Thus, any PMS satisfying $c^T_{i} < d^T_{i}+ b^T_{i}$ is strictly dominated, as shown in Lemma~\ref{one-horizon}.

By eliminating such strictly dominated strategies, we define the subset of pure PMS for user $i$ as $\mathcal{\bar P}^{PMS}_{i}:= \{\boldsymbol{\mathbbm{1}}_{\boldsymbol{a}_{i}} \in \mathcal{P}^{PMS}_{i}\mid \forall \boldsymbol s^{T} \in \mathcal{S}^{T}, c^T_{i} = d^T_{i}+ b^T_{i} \}$. Clearly, $\mathcal{\bar P}^{PMS}_{i} \subsetneqq \mathcal{P}^{PMS}_{i}$.  

Since the elimination of strictly dominated strategies does not affect equilibrium computation, we substitute $r_i(\boldsymbol{s}^T,\boldsymbol{a}^T) = \theta_i  c_i^T - P(d^{T}, e^T_{N})  d_i^T$ and $c^T_{i} = d^T_{i}+ b^T_{i}$ into \eqref{Expanding the expected value-function}, yielding:
\begin{align}\label{removing strictly dominated strategies}
     &\max \limits_{\boldsymbol{\mathbbm{1}}_{\boldsymbol{a}_{i}} \in \mathcal{P}^{PMS}_{i}} V^{\boldsymbol{\mathbbm{1}}_{\boldsymbol{a}_{i}},\boldsymbol{\mathbbm{1}}_{\boldsymbol{a}_{-i}}}_{1,i}(\boldsymbol s^1)  =\max\limits_{\boldsymbol{\mathbbm{1}}_{\boldsymbol{a}_{i}} \in \mathcal{\bar P}^{PMS}_{i}}V^{\boldsymbol{\mathbbm{1}}_{\boldsymbol{a}_{i}},\boldsymbol{\mathbbm{1}}_{\boldsymbol{a}_{-i}}}_{1,i}(\boldsymbol s^1) 
 \notag \\                                                &=\max\limits_{\boldsymbol{\mathbbm{1}}_{\boldsymbol{a}_{i}} \in \mathcal{\bar P}^{PMS}_{i}} \Bigg\{r_{i}\left(\boldsymbol s^{1}, \boldsymbol a^{1}\right) + \sum_{\boldsymbol s^{2} \in \mathcal{S}^2}q(\boldsymbol s^{2} \mid  \boldsymbol s^1, \boldsymbol a^{1}) \Big[ r_{i}\left(\boldsymbol s^{2}, \boldsymbol a^{2}\right) +\cdots \notag \\   &\quad +   \sum_{\boldsymbol s^{T-1} \in \mathcal{S}^{T-1}}q(\boldsymbol s^{T-1} \mid  \boldsymbol s^{T-2}, \boldsymbol a^{T-2}) \big[r_{i}\left(\boldsymbol s^{T-1}, \boldsymbol a^{T-1}\right) + \sum_{\boldsymbol s^{T} \in \mathcal{S}^{T}}q(\boldsymbol s^{T} \mid  \boldsymbol s^{T-1}, \boldsymbol a^{T-1})\times\notag \\
    & \quad\quad\underbrace{[ 
    g_i(e^T_{N},\boldsymbol{d}^T) + \theta_i  b^T_i ]}_{(T) *}\big]\cdots\Big]\Bigg\},
\end{align} 
where $g_i(e^T_{N},\boldsymbol{d}^T) := \theta_i  d^T_{i} - P(d^{T}, e^T_{N})  d^{T}_{i}$, as defined in \eqref{max_g}, represents the stage payoff determined solely by electricity demand, and constitutes part of the final-stage payoff $(T)*$.

Now, given $\boldsymbol s^{T} \in\mathcal{S}^{T}$, consider the corresponding single-stage game, in which user~$i$'s action space is the reduced set $\mathcal{\bar A}^T_i(\boldsymbol s^{T})$ \eqref{single-A}, and the payoff is the term $(T) *$ in \eqref{removing strictly dominated strategies}. By Lemma~\ref{one-horizon}, this stage game admits a demand equilibrium $\boldsymbol{d}^{*T}(e^T_{N})$ that depends only on the $e^T_{N}$. 
Moreover, for any user $i$, any strategy outside the set $\mathcal{\bar P}^{PMS}_i$ is strictly dominated. Therefore, after eliminating such strategies, we obtain the equilibrium of the stage-$T$ game, denoted by $(d_i^{*T}(e^T_{N}), c_i^{*T}(\boldsymbol s_i^{T}))$ with $c_i^{*T}(\boldsymbol s_i^{T}) = d_i^{*T}(e^T_{N}) + b^T_i, \forall i \in \mathcal{N}$. It is reasonable to assume that all users adopt this equilibrium strategy at stage $T$. Thus, substituting \(\boldsymbol{d}^{*T}(e^T_{N})\) into~\eqref{removing strictly dominated strategies}, user~\(i\) only needs to focus on strategy selection over the first \(T{-}1\) stages to maximize their payoff. Accordingly, the objective in~\eqref{removing strictly dominated strategies} reduces to:
\begin{align}\label{Substituting the equilibrium}
    &\max\limits_{\boldsymbol{\mathbbm{1}}_{\boldsymbol{a}_{i}} \in \mathcal{\bar P}_{i}^{T-1}}V^{\boldsymbol{\mathbbm{1}}_{\boldsymbol{a}_{i}},\boldsymbol{\mathbbm{1}}_{\boldsymbol{a}_{-i}}}_{1,i}(\boldsymbol s^1)  \notag \\       &=\max\limits_{\boldsymbol{\mathbbm{1}}_{\boldsymbol{a}_{i}} \in \mathcal{\bar P}_{i}^{T-1}} \Bigg\{r_{i}\left(\boldsymbol s^{1}, \boldsymbol a^{1}\right) + \sum_{\boldsymbol s^{2} \in \mathcal{S}^2}q(\boldsymbol s^{2} \mid  \boldsymbol s^1, \boldsymbol a^{1}) \Big[ r_{i}\left(\boldsymbol s^{2}, \boldsymbol a^{2}\right)  +\cdots  \notag \\
    & \quad +   \sum_{\boldsymbol s^{T-1} \in \mathcal{S}^{T-1}}q(\boldsymbol s^{T-1} \mid  \boldsymbol s^{T-2}, \boldsymbol a^{T-2}) \big[r_{i}\left(\boldsymbol s^{T-1}, \boldsymbol a^{T-1}\right) + \sum_{\boldsymbol s^{T} \in \mathcal{S}^{T}}q(\boldsymbol s^{T} \mid  \boldsymbol s^{T-1}, \boldsymbol a^{T-1}) \times\notag \\
    & \quad\quad [ 
    g_i(e^T_{N},\boldsymbol{d}^{*T}) + \theta_i  b^T_i ]\big]\cdots\Big]\Bigg\},
\end{align}
where, $\mathcal{\bar P}_{i}^{T-1}:= \{\boldsymbol{\mathbbm{1}}_{\boldsymbol{a}_{i}} \in \mathcal{\bar P}^{PMS}_{i}\mid \forall \boldsymbol s^{T} \in \mathcal{S}^{T},\mathbbm{1}_{a_i^{t}}(\boldsymbol s^T_i) = (d^{*T}_i(e^{T}_{N}), d^{*T}_i(e^{T}_{N}) + b_i^T)\}$ and we assume \(\boldsymbol{\mathbbm{1}}_{\boldsymbol{a}_{-i}} \in \mathcal{\bar P}_{-i}^{T-1}\), where \(\mathcal{\bar P}_{-i}^{T-1}\) is defined analogously. To generalize this reduced strategy set for earlier stages, we define: 
\begin{align*}
 \mathcal{\bar P}_{i}^{t}:= \{\boldsymbol{\mathbbm{1}}_{\boldsymbol{a}_{i}} \in \mathcal{\bar P}^{PMS}_{i}\mid \forall t^{\prime} \in \{t+1,\dots,T\}, \boldsymbol s^{t^{\prime}} \in \mathcal{S}^{t^{\prime}}
 \mathbbm{1}_{a_i^{t}}(\boldsymbol s^{t^{\prime}}_i)
= (d^{*t}_i(e^{t^{\prime}}_{N}), d^{*t}_i(e^{t^{\prime}}_{N}) + b_i^{t^{\prime}})\},
\end{align*}
which emphasizes user $i$ adopts the equilibrium strategy at every future stage $t^{\prime} > t$.

\textit{Step 2. Reformulation of expected value at Stage $T$:}
Next, consider the term $\sum_{\boldsymbol s^{T} \in \mathcal{S}^{T}}q(\boldsymbol s^{T} \mid  \boldsymbol s^{T-1}, \boldsymbol a^{T-1})\times $ $[ 
    g_i(e^T_{N},\boldsymbol{d}^{*T}) + \theta_i  b^T_i ]$ in \eqref{Substituting the equilibrium}. Since $g_i(e^T_{N}, \boldsymbol{d}^{*T})$ depends only on the public state component $e^T_{N}$, and the transition probability of $e^T_{N}$ follows $\hat q(e^{T}_{N} \mid e^{T-1}_{N})$, which is independent of both $\boldsymbol a^{T-1}$ and $b_i^{T-1}$ for any $i \in \mathcal{N}$, we obtain:
\begin{align}\label{transition simplify}
     &\sum_{\boldsymbol s^{T} \in \mathcal{S}^{T}}q(\boldsymbol s^{T} \mid  \boldsymbol s^{T-1}, \boldsymbol a^{T-1}) [g_i(e^T_{N}, \boldsymbol{d}^{*T})] \notag \\
     &= \sum_{e^T_{N} \in \mathcal E^T_{N}}  \widetilde q(e^{T}_{N} \mid  \boldsymbol s^{T-1}, \boldsymbol a^{T-1}) [g_i(e^T_{N},\boldsymbol{d}^{*T})] \notag \\
     & = \sum_{e^T_{N} \in \mathcal E^T_{N}}  \hat  q(e^{T}_{N} \mid  e^{T-1}_{N}) [g_i(e^T_{N}, \boldsymbol{d}^{*T})], 
\end{align}
where $\widetilde q(e^{t}_{N} \mid  \boldsymbol s^{t-1}, \boldsymbol a^{t-1})$ denotes the marginal distribution of $e^t_N$ obtained from $q(\boldsymbol s^{t} \mid \boldsymbol s^{t-1}, \boldsymbol a^{t-1})$ by marginalizing over the storage level $(b^t_i)_{i \in \mathcal{N}}$. 

Moreover, given $\boldsymbol s^{T-1}, \boldsymbol a^{T-1}$, the storage level update rule $b^T_i = b^{T-1}_i + d^{T-1}_i - c^{T-1}_i$ for all $i \in \mathcal{N}$ implies:
\begin{align}\label{storage state simplify}
     &\sum_{\boldsymbol s^{T} \in \mathcal{S}^{T}}q(\boldsymbol s^{T} \mid  \boldsymbol s^{T-1}, \boldsymbol a^{T-1}) [\theta_i  b^T_i ] \notag \\
     &= \sum_{\boldsymbol s^{T} \in \mathcal{S}^{T}}q(\boldsymbol s^{T} \mid  \boldsymbol s^{T-1}, \boldsymbol a^{T-1})[\theta_i  (b^{T-1}_i + d^{T-1}_i-c^{T-1}_i) ] \notag \\
     &= \theta_i  (b^{T-1}_i + d^{T-1}_i-c^{T-1}_i).
\end{align}

Substituting \eqref{storage state simplify} and \eqref{transition simplify} into \eqref{Substituting the equilibrium} , we can obtain  a reformulated expression for the expected value at stage~$T$, and rewrite~\eqref{Substituting the equilibrium} as follows: 
\begin{align}\label{T stage simplify}
    &\max\limits_{\boldsymbol{\mathbbm{1}}_{\boldsymbol{a}_{i}} \in \mathcal{\bar P}_{i}^{T-1}}V^{\boldsymbol{\mathbbm{1}}_{\boldsymbol{a}_{i}},\boldsymbol{\mathbbm{1}}_{\boldsymbol{a}_{-i}}}_{1,i}(\boldsymbol s^1)  \notag \\  
    &= \max\limits_{\boldsymbol{\mathbbm{1}}_{\boldsymbol{a}_{i}} \in \mathcal{\bar P}_{i}^{T-1}} \Bigg\{r_{i}\left(\boldsymbol s^{1}, \boldsymbol a^{1}\right) + \sum_{\boldsymbol s^{2} \in \mathcal{S}^2}q(\boldsymbol s^{2} \mid  \boldsymbol s^1, \boldsymbol a^{1}) \Big[ r_{i}\left(\boldsymbol s^{2}, \boldsymbol a^{2}\right)  \notag \\ &\quad+\cdots +   \sum_{\boldsymbol s^{T-1} \in \mathcal{S}^{T-1}}q(\boldsymbol s^{T-1} \mid  \boldsymbol s^{T-2}, \boldsymbol a^{T-2}) \big[r_{i}\left(\boldsymbol s^{T-1}, \boldsymbol a^{T-1}\right) + \theta_i  (b^{T-1}_i + d^{T-1}_i-c^{T-1}_i) \notag \\ &\quad+ \sum_{e^T_{N} \in \mathcal E^T_{N}}  \hat  q(e^{T}_{N} \mid  e^{T-1}_{N}) [g_i(e^T_{N}, \boldsymbol{d}^{*T})]  \big]\cdots\Big]\Bigg\}.
\end{align}

\textit{Step 3. Recursive Equation at stage $T-1$:} In \eqref{T stage simplify}, the stage payoff at stage $T-1$ is $r_i(\boldsymbol{s}^{T-1},\boldsymbol{a}^{T-1}) = \theta_i  c_i^{T-1} - P(d^{T-1}, e^{T-1}_{N})  d_i^{T-1}$. Substitute it to \eqref{T stage simplify} and we have:
\begin{align}\label{T-1 expanding}
     &\max\limits_{\boldsymbol{\mathbbm{1}}_{\boldsymbol{a}_{i}} \in \mathcal{\bar P}_{i}^{T-1}}V^{\boldsymbol{\mathbbm{1}}_{\boldsymbol{a}_{i}},\boldsymbol{\mathbbm{1}}_{\boldsymbol{a}_{-i}}}_{1,i}(\boldsymbol s^1) \notag \\
     & =\max\limits_{\boldsymbol{\mathbbm{1}}_{\boldsymbol{a}_{i}} \in \mathcal{\bar P}_{i}^{T-1}} \Bigg\{r_{i}\left(\boldsymbol s^{1}, \boldsymbol a^{1}\right) + \sum_{\boldsymbol s^{2} \in \mathcal{S}^2}q(\boldsymbol s^{2} \mid  \boldsymbol s^1, \boldsymbol a^{1})  \Big[ r_{i}\left(\boldsymbol s^{2}, \boldsymbol a^{2}\right)  \notag \\ &\quad +\cdots + \sum_{\boldsymbol s^{T-1} \in \mathcal{S}^{T-1}}q(\boldsymbol s^{T-1} \mid  \boldsymbol s^{T-2}, \boldsymbol a^{T-2})\big[  g_i(e^{T-1}_{N},\boldsymbol{d}^{T-1}) + \theta_i  b^{T-1}_i \notag \\ &\quad + \sum_{e^T_{N} \in \mathcal E^T_{N}} \hat  q(e^{T}_{N} \mid  e^{T-1}_{N}) [ 
    g_i(e^T_{N}, \boldsymbol{d}^{*T})] \big]\cdots\Big]\Bigg\} \notag \\
    & =\max\limits_{\boldsymbol{\mathbbm{1}}_{\boldsymbol{a}_{i}} \in \mathcal{\bar P}_{i}^{T-1}} \Bigg\{r_{i}\left(\boldsymbol s^{1}, \boldsymbol a^{1}\right) + \sum_{\boldsymbol s^{2} \in \mathcal{S}^2}q(\boldsymbol s^{2} \mid  \boldsymbol s^1, \boldsymbol a^{1}) \Big[ r_{i}\left(\boldsymbol s^{2}, \boldsymbol a^{2}\right) \notag \\ 
    &\quad + \cdots  + \sum_{\boldsymbol s^{T-1} \in \mathcal{S}^{T-1}}q(\boldsymbol s^{T-1} \mid  \boldsymbol s^{T-2}, \boldsymbol a^{T-2}) \big[  g_i(e^{T-1}_{N},\boldsymbol{d}^{T-1}) + \theta_i  b^{T-1}_i\big] \notag \\
    & \quad + \underbrace{\sum_{\boldsymbol s^{T-1} \in \mathcal{S}^{T-1}}q(\boldsymbol s^{T-1} \mid  \boldsymbol s^{T-2}, \boldsymbol a^{T-2})[\sum_{e^T_{N} \in \mathcal E^T_{N}} \hat  q(e^{T}_{N} \mid  e^{T-1}_{N}) 
    g_i(e^T_{N}, \boldsymbol{d}^{*T})]}_{\circledast}\big]\cdots\Big]\Bigg\}.
\end{align} 
Next, consider the term $\circledast$ in \eqref{T-1 expanding}. In $\circledast$, as in stage $T$, the quantity $\sum_{e^T_{N} \in \mathcal E^T_{N}} \hat q(e^{T}_{N} \mid e^{T-1}_{N})  g_i(e^T_{N}, \boldsymbol{d}^{*T})$ depends only on the public state $e^{T-1}_{N}$, and the transition probability $\hat q(e^{T-1}_{N} \mid e^{T-2}_{N})$ is independent of both $\boldsymbol a^{T-2}$ and $b_i^{T-2}$ for any $i \in \mathcal{N}$. Hence, similar to \eqref{transition simplify}, we rewrite :
\begin{align*}
     &\sum_{\boldsymbol s^{T-1} \in \mathcal{S}^{T-1}}q(\boldsymbol s^{T-1} \mid  \boldsymbol s^{T-2}, \boldsymbol a^{T-2}) [\sum_{e^T_{N} \in \mathcal E^T_{N}} \hat  q(e^{T}_{N} \mid  e^{T-1}_{N})   
    g_i(e^T_{N}, \boldsymbol{d}^{*T})]  \\
     & = \sum_{e^{T-1}_{N} \in \mathcal E^{T-1}_{N}} \hat  q(e^{T-1}_{N} \mid  e^{T-2}_{N}) [\sum_{e^T_{N} \in \mathcal E^T_{N}} \hat  q(e^{T}_{N} \mid  e^{T-1}_{N})  
    g_i(e^T_{N}, \boldsymbol{d}^{*T})]. 
\end{align*}
 
Substitute this result into \eqref{T-1 expanding} yields the recursive equation at stage $T-1$:
 \begin{align}\label{T-1 final payoff}
     &\max\limits_{\boldsymbol{\mathbbm{1}}_{\boldsymbol{a}_{i}} \in \mathcal{\bar P}_{i}^{T-1}}V^{\boldsymbol{\mathbbm{1}}_{\boldsymbol{a}_{i}},\boldsymbol{\mathbbm{1}}_{\boldsymbol{a}_{-i}}}_{1,i}(\boldsymbol s^1)  := \max\limits_{\boldsymbol{\mathbbm{1}}_{\boldsymbol{a}_{i}} \in \mathcal{\bar P}_{i}^{T-1}} \Bigg\{r_{i}\left(\boldsymbol s^{1}, \boldsymbol a^{1}\right) +\sum_{\boldsymbol s^{2} \in \mathcal{S}^2}q(\boldsymbol s^{2} \mid  \boldsymbol s^1, \boldsymbol a^{1}) \Big[ r_{i}\left(\boldsymbol s^{2}, \boldsymbol a^{2}\right) \notag \\
     & \quad + \cdots + \sum_{\boldsymbol s^{T-1} \in \mathcal{S}^{T-1}}q(\boldsymbol s^{T-1} \mid  \boldsymbol s^{T-2}, \boldsymbol a^{T-2})  \underbrace{\big[  g_i(e^{T-1}_{N},\boldsymbol{d}^{T-1}) + \theta_i  b^{T-1}_i \big]}_{(T-1) *} \notag \\                               &\quad + \mathbf p^{T-2}_{T-1}\mathbf Q^{T-1}_{T}\mathbf g_i^T(\boldsymbol{d}^{*T})\big]\cdots\Big]\Bigg\},   \end{align} 
where  
\begin{align*}  
\mathbf {p}^{t}_{t+1} &= \big[ \hat{q}(e^{t+1}_{N} \mid e^t_{N}) \big]^{\top}_{e^{t+1}_{N} \in \mathcal{E}^{t+1}_{N}}, \notag \\  
\mathbf Q^{t+1}_{t+2} &= \big[ \hat{q}(e^{t+2}_{N} \mid e^{t+1}_{N}) \big]_{e^{t+1}_{N}\in \mathcal{E}^{t+1}_{N}, e^{t+2}_{N} \in \mathcal{E}^{t+2}_{N}}, \notag \\  
\mathbf g_i^t(\boldsymbol{d}^{*t}) &= \big[ g_i(e^t_{N}, \boldsymbol{d}^{*t}) \big]_{e^t_{N} \in \mathcal{E}^{t}_{N}}.  
\end{align*}  
Intuitively, $\mathbf {p}^{t}_{t+1}$ is the row vector of transition probabilities from stage $t$ to stage $t+1$, $\mathbf Q^{t+1}_{t+2}$ is the transition probability matrix from stage $t+1$ to stage $t+2$, and $\mathbf g_i^t(\boldsymbol{d}^{*t})$ is the column vector of demand equilibrium reward $g_i(e^t_{N}, \boldsymbol{d}^{*t})$.

\textit{Step 4. Backward Induction for All Stages $t < T$:} In \eqref{T-1 final payoff}, the stage payoff term $(T-1) *$ has a similar form to the term $(T) *$ in \eqref{Expanding the expected value-function}. Therefore, we can apply backward induction by recursively repeating Step 1-3 for all stages $t < T$, ultimately reaching stage~$t = 1$:
\begin{align}\label{max_t=1}
     &\max\limits_{\boldsymbol{\mathbbm{1}}_{\boldsymbol{a}_{i}} \in \mathcal{\bar P}_{i}^{1}}V^{\boldsymbol{\mathbbm{1}}_{\boldsymbol{a}_{i}},\boldsymbol{\mathbbm{1}}_{\boldsymbol{a}_{-i}}}_{1,i}(\boldsymbol s^1)\notag \\
&=\max\limits_{\boldsymbol{\mathbbm{1}}_{\boldsymbol{a}_{i}} \in \mathcal{\bar P}_{i}^{1}} \Bigg\{ r_{i}\left(\boldsymbol s^{1}, \boldsymbol a^{1}\right) + \theta_i  (b^{1}_i + d^{1}_i-c^{1}_i) + \mathbf p^{1}_{2} \mathbf g_i^2(\boldsymbol{d}^{*2})\notag \\
    & \quad + \mathbf p^{1}_{2}\mathbf Q ^{2}_{3}\mathbf g_i^3(\boldsymbol{d}^{*3}) + \cdots + \mathbf p^{1}_{2}\mathbf Q^{2}_{3}...\mathbf Q^{T-1}_{T}\mathbf g_i^T(\boldsymbol{d}^{*T})\Bigg\} \notag \\
    & =\max\limits_{\boldsymbol{\mathbbm{1}}_{\boldsymbol{a}_{i}} \in \mathcal{\bar P}_{i}^{1}} \Bigg\{ \underbrace{g_i(e^{1}_{N},\boldsymbol{d}^{1}) + \theta_i  b^{1}_i }_{(1) *} +  \mathbf p^{1}_{2} \mathbf g_i^2(\boldsymbol{d}^{*2}) + \mathbf p^{1}_{2}\mathbf Q ^{2}_{3}\mathbf g_i^3(\boldsymbol{d}^{*3}) \notag \\
    & \quad + \cdots  + \mathbf p^{1}_{2}\mathbf Q^{2}_{3}\dots \mathbf Q^{T-1}_{T}\mathbf g_i^T(\boldsymbol{d}^{*T})  \Bigg\}.
\end{align}
In \eqref{max_t=1}, the remaining terms are the expected equilibrium payoffs for all future stages and can thus be regarded as constants. Then we only need to consider the payoff term $(1) *$, which effectively reduces the problem to a stage game where user~$i$’s action is the demand $d_i^1$, and the stage payoff is given by $(1) *$. By Lemma~\ref{one-horizon}, a stage-wise equilibrium $\boldsymbol{d}^{*1}(e^{1}_{N})$ exists.

By combining the stage-wise equilibrium strategies for all stages, we can construct a pure PMS profile. Moreover, based on the above construction and the principle of backward induction, this pure PMS profile constitutes a pure PME:
$(\boldsymbol{\mathbbm{1}}_{\boldsymbol{a}^{*}_{i}} ,\boldsymbol{\mathbbm{1}}_{\boldsymbol{a}^{*}_{-i}})$, where for all $i \in \mathcal{N}, t \in \mathcal{T}, \boldsymbol s_i^t \in \mathcal{S}_i^t$,
\begin{equation}\label{case2}
\mathbbm{1}_{\boldsymbol{a}^{*t}_{i}} (\boldsymbol{s}_i^t) =(d^{*t}_i(e^{t}_{N}),c^{*t}_i(\boldsymbol s^t_i)) = (d^{*t}_i(e^{t}_{N}),d^{*t}_i(e^{t}_{N}) + b_i^t) .
\end{equation}
    
Under this pure PME, the corresponding value function for user $i$ is given by:
\begin{align}
V^{\boldsymbol{\mathbbm{1}}_{\boldsymbol{a}^{*}_{i}},\boldsymbol{\mathbbm{1}}_{\boldsymbol{a}^{*}_{-i}}}_{1,i}(\boldsymbol s^1)
    &= g_i(e^1_{N}, \boldsymbol{d}^{*1}) +  \mathbf p^{1}_{2} \mathbf g_i^2(\boldsymbol{d}^{*2}) + \mathbf p^{1}_{2}\mathbf Q ^{2}_{3}\mathbf g_i^3(\boldsymbol{d}^{*3})  + \cdots  + \mathbf p^{1}_{2}\mathbf Q^{2}_{3}\dots \mathbf Q^{T-1}_{T}\mathbf g_i^T(\boldsymbol{d}^{*T}). \label{eq:value_function} 
\end{align}
This completes the proof of the theorem.
\te

\section{Coefficients List and Pure Equilibrium Demand in Example~\ref{FIP example}}\label{appendix4}
\[
\{\theta_i\}_{i=1}^{50} =
\begin{bmatrix}
1.019 & 1.01 & 1.021 & 1.025 & 1.002 & 1.02 & 1.2 & 1.3 & 1.4 & 1.5 \\
0.9   & 1    & 1.1   & 1.15  & 1.32  & 1.22 & 1.23 & 1.33 & 1.34 & 1.35 \\
0.9   & 1.1  & 1.01  & 1.05  & 1.12  & 1.02 & 1.12 & 1.03 & 1.04 & 1.05 \\
0.9   & 1    & 1.01  & 1.05  & 1.042 & 1.032& 1.012& 1.023& 1.014& 1.025 \\
1.019 & 1    & 1.01  & 1.05  & 1.02  & 1.02 & 1.12 & 1.13 & 1.14 & 1.01 \\
\end{bmatrix}.
\]

\[
\begin{array}{c@{\quad\quad}c}
\{\boldsymbol{\mathbbm{1}}_{\boldsymbol{d}^{\ast}_{i}}(\cdot \mid t=3,e_{N}^3 = 90)\}_{i=1}^{50}= &
\{\boldsymbol{\mathbbm{1}}_{\boldsymbol{d}^{\ast}_{i}}(\cdot \mid t=1,e_{N}^1 = 70)\}_{i=1}^{50}= \\
\begin{bmatrix}
4 & 4 & 4 & 4 & 3 & 4 & 4 & 4 & 4 & 4 \\
0 & 2 & 4 & 4 & 4 & 4 & 4 & 4 & 4 & 4 \\
0 & 4 & 4 & 4 & 4 & 4 & 4 & 4 & 4 & 4 \\
0 & 2 & 4 & 4 & 4 & 4 & 4 & 4 & 4 & 4 \\
4 & 3 & 4 & 4 & 4 & 4 & 4 & 4 & 4 & 4 \\
\end{bmatrix}
&
\begin{bmatrix}
2 & 1 & 2 & 3 & 0 & 2 & 4 & 4 & 4 & 4 \\
0 & 0 & 4 & 4 & 4 & 4 & 4 & 4 & 4 & 4 \\
0 & 4 & 1 & 4 & 4 & 2 & 4 & 4 & 4 & 4 \\
0 & 0 & 1 & 4 & 4 & 4 & 1 & 3 & 1 & 3 \\
2 & 0 & 1 & 4 & 3 & 2 & 4 & 4 & 4 & 0 \\
\end{bmatrix}
\end{array}
\]

\section{Details of MDP and Algorithm~\ref{MDP} in \ref{second alg}}\label{appendix5}
\begin{align*}
    M_{i}^{k}:=\{\mathcal{T},\{\mathcal{S}_i^t\}_{t=1}^T,\mathcal A_{i},r_{i}^{k}, q_{i}^{k}\},
\end{align*}
where
\begin{align*}
&r^k_i(e^t_{N}, b^t_i, a^t_i)  :=  \theta_{i} c_{i}^{t} - P^k(d^{t}, e^t_{N}) d^{t}_{i},
\end{align*}
\begin{align*}
 &P^k(d^{t}, e^t_{N}) := \sum_{d_{-i}^{t} \in \mathcal D_{-i}} \hat{\boldsymbol \pi}_{-i}^{k}(  d_{-i}^{t} \mid e_{N}^{t})
\left[ \frac{\alpha}{n e_{N}^{t} + \gamma_{1}}   (d_{-i}^{t} + d_{i}^{t}) + \frac{\beta}{e_{N}^{t} + \gamma_{2}} \right],
\end{align*}
\begin{align*}
q^k_i(e^{t+1}_{N} = \hat e^{t+1}_{N} + \omega^{t+1}_{N}, b^{t+1}_i) \mid e^t_{N} = \hat e^t_{N} +& \omega^t_{N}, b^t_i, a^t_i)   \\
:= \begin{cases}
\hat q(\omega^{t+1}_{N} \mid \omega^{t}_{N}),  &\text{if } b^{t+1}_{i}=b^{t}_{i}+d^{t}_{i}-c^{t}_{i}, \\
0,  &\text{otherwise}.
\end{cases}
\end{align*}

\begin{algorithm}[H]
\caption{Backward Induction for T-Horizon MDP \cite{puterman2014markov}.}
\label{MDP}
\begin{algorithmic}
\STATE \textbf{Input:} $\mathcal{T}, \mathcal{S}, \mathcal{A}, \{P^t\}_{t \in \mathcal{T}}, \{r^t\}_{t \in \mathcal{T}}$.
\STATE \textbf{Initialization:}
\STATE \hspace{0.5cm} Set $t = T$, $V(T, s) = r^T(s), \forall s \in \mathcal{S}$. 
\STATE \hspace{0.5cm} Initialize strategy $\boldsymbol \pi: \mathcal{T} \times \mathcal{S} \rightarrow \Delta(\mathcal{A})$.
\FOR{$t = T-1$ \TO $1$}
    \FOR{each state $s \in \mathcal{S}$}
\STATE $V(t,s) = \max_{a \in \mathcal{A}} [ r^t(s,a) + \sum_{s'} P^t(s'|s,a) V(t+1,s') ].$
        \STATE Set $\boldsymbol \pi(t,s)$ to the maximizing action.
    \ENDFOR
\ENDFOR
\STATE \textbf{Output:} Optimal strategy $\boldsymbol \pi(t, s)$ for all $t \in \mathcal{T}, s \in \mathcal{S}$.
\end{algorithmic}
\end{algorithm}

\section{Verification of the Strict Dominance of Energy Storage Strategies}\label{appendix3}

For each $i =1,2$, to verify the energy storage strategy 
$\boldsymbol{\mathbbm{1}}_{\boldsymbol{ \bar a}_{i}} :=\{ \mathbbm{1}_{{ \bar a}^1_{i}} (e_{N}^1=10,b_i^1=0) =(\bar d_i^1=3,\bar c_i^1=2);\mathbbm{1}_{{ \bar a}^2_{i}} (e_{N}^2,b_i^2) =(\bar d_i^2=1,\bar c_i^2=2),\forall e_{N}^2 \in \mathcal E^2_{N}, b_i^2 \in \mathcal{B}_{i}\}$ strictly dominates any non-storage strategy $\boldsymbol{\mathbbm{1}}_{\hat{\boldsymbol{a}}_{i}} \in \mathcal{P}_{i}^{\boldsymbol d,T} := \{\boldsymbol{\mathbbm{1}}_{\boldsymbol{a}_{i}} \in \mathcal{P}^{PMS}_{i}\mid c^t_i = d^t_i \in \{1,2,3\} ,\forall t =1,2 \}$, we first consider the temporal action pair $\boldsymbol{d}_{i}^{(1,2)}:= (d_i^1,d_i^2)$, which represents the user $i$'s  electricity
demand at two time steps. We denote $\boldsymbol{d}_{i}^{(1,2)} \in \mathcal D_{i}^{(1,2)}:= \mathcal D_{i} \times \mathcal D_{i} = [1,2,3]\times [1,2,3]$ and $\boldsymbol{d}^{(1,2)} := (\boldsymbol{d}_{i}^{(1,2)} ,\boldsymbol{d}_{-i}^{(1,2)} )$ as the joint temporal action pair of all users. The instantaneous reward of user $i$, given the joint temporal action pair $\boldsymbol{d}^{(1,2)}$, is defined as the sum of rewards over the two time steps, considering any possible states pair $(e^1_{N},e^2_{N}) \in \mathcal E_{N}^1 \times \mathcal E_{N}^2$ without specifying state transitions:
\begin{align*}
r_i(d^1_i, d^1_{-i})_{e^1_{N}} + r_i(d^2_i, d^2_{-i})_{e^2_{N}}
= H_i(d_i^1) - P(d^{1}, e^1_{N}) \cdot d^{1}_{i} + H_i(d_i^2) - P(d^{2}, e^2_{N}) \cdot d^{2}_{i}.
\end{align*}

Similarly, for the energy storage strategy $\boldsymbol{\mathbbm{1}}_{\boldsymbol{ \bar a}_{i}}$, we define the corresponding temporal action pair
$\boldsymbol{\bar a}_{i}^{(1,2)}:= (\bar c_i^1,\bar d_i^1,\bar c_i^2,\bar d_i^2)$ and the instantaneous reward of user $i$ under the joint temporal action pair $\boldsymbol{\bar a}^{(1,2)} := (\boldsymbol{\bar a}_{i}^{(1,2)} ,\boldsymbol{a}_{-i}^{(1,2)} )$ as:
\begin{align*}
& r_i(\bar d^1_i,\bar c^1_i, d^1_{-i},c^1_{-i})_{e^1_{N}} + r_i(\bar d^2_i, \bar c^2_i, d^2_{-i}, c^2_{-i})_{e^2_{N}} \\& = H_i(\bar c_{i}^1) - P( d^{1}, e^1_{N}) \cdot \bar d^{1}_{i} + H_i(\bar c_{i}^2)  - P(d^{2}, e^2_{N}) \cdot \bar d^{2}_{i}.
\end{align*}

From the structure of the instantaneous reward, we can verify that if the following condition holds:
\begin{align*}
\text{C1.} \frac{\alpha}{n e^1_{N}+\gamma_{1}}\cdot (5+d^{1}_{-i})+\frac{\beta} {e^1_{N}+\gamma_{2}} < 0.9
< \frac{\alpha}{n e^2_{N}+\gamma_{1}}\cdot (3+d^{2}_{-i})+\frac{\beta} {e^2_{N}+\gamma_{2}} ,\\ \forall e_{N}^1 \in \mathcal E^1_{N}, e_{N}^2 \in \mathcal E^2_{N},d^{1}_{-i},d^{2}_{-i} \in \mathcal D_{-i}=\mathcal D_{i}, 
\end{align*}
then the instantaneous reward associated with the temporal action pair of the energy storage strategy $\boldsymbol{\mathbbm{1}}_{\boldsymbol{ \bar a}_{i}}$ strictly dominates that of any non-storage strategy $\boldsymbol{\mathbbm{1}}_{\boldsymbol{d}_{i}} \in \mathcal{P}_{i}^{\boldsymbol d,T}$ for user $i$, i.e., 
\begin{align*}
r_i(\bar d^1_i,\bar c^1_i, d^1_{-i},c^1_{-i})_{e^1_{N}} + r_i(\bar d^2_i, \bar c^2_i, d^2_{-i}, c^2_{-i})_{e^2_{N}}  > r_i(d^1_i, d^1_{-i})_{e^1_{N}} + r_i(d^2_i, d^2_{-i})_{e^2_{N}} ,\\ \forall e_{N}^1 \in \mathcal E^1_{N}, e_{N}^2 \in \mathcal E^2_{N},d^{1}_{-i},d^{2}_{-i} \in \mathcal D_{-i}=\mathcal D_{i}. \notag
\end{align*}
Clearly, setting $\alpha = 1, \beta =1.5, \gamma_1=\gamma_2 = 1$ satisfies condition C1.

Therefore, given that condition C1 holds and considering the linear relationship between the value function $V^{\boldsymbol{\mathbbm{1}}_{\boldsymbol{a}_{i}},\boldsymbol{\mathbbm{1}}_{\boldsymbol{a}_{-i}}}_{1,i}(\boldsymbol s^{1})$ and the instantaneous reward, we conclude that the energy storage strategy $\boldsymbol{\mathbbm{1}}_{\boldsymbol{ \bar a}_{i}}$ strictly dominates any non-storage strategy $\boldsymbol{\mathbbm{1}}_{\boldsymbol{d}_{i}} \in \mathcal{P}_{i}^{\boldsymbol d,T}$, i.e.,
\begin{align*} 
&V^{\boldsymbol{\mathbbm{1}}_{\boldsymbol{\bar a}_{i}},\boldsymbol{\mathbbm{1}}_{\boldsymbol{a}_{-i}}}_{1,i}(\boldsymbol s^{1}) \\
&= r_i(\bar d^1_i,\bar c^1_i, d^1_{-i},c^1_{-i})_{e^1_{N=10}} + 0.3 \cdot r_i(\bar d^2_i, \bar c^2_i, d^2_{-i}, c^2_{-i})_{e^2_{N=1}}  \\
&\quad + 0.4 \cdot r_i(\bar d^2_i, \bar c^2_i, d^2_{-i}, c^2_{-i})_{e^2_{N=2}}  +0.3 \cdot r_i(\bar d^2_i, \bar c^2_i, d^2_{-i}, c^2_{-i})_{e^2_{N=3}} \\  
& > \hat V^{\boldsymbol{\mathbbm{1}}_{\boldsymbol{d}_{i}},\boldsymbol{\mathbbm{1}}_{\boldsymbol{d}_{-i}}}_{1,i}(\boldsymbol s^1)  \\
& \quad = r_i(d^1_i, d^1_{-i})_{e^1_{N=10}} + 0.3 \cdot r_i(d^2_i, d^2_{-i})_{e^2_{N=1}}  \\  &\quad\quad + 0.4 \cdot r_i(d^2_i, d^2_{-i})_{e^2_{N=2}}  +0.3 \cdot r_i(d^2_i, d^2_{-i})_{e^2_{N=3}}, \\ 
&\quad\quad \forall \boldsymbol{\mathbbm{1}}_{\boldsymbol{a}_{-i}} \in \mathcal{P}_{-i}  \text{ and corresponding } \boldsymbol{\mathbbm{1}}_{\boldsymbol{d}{-i}} \in \mathcal{P}_{-i}^{\boldsymbol{d},T}, \boldsymbol{\mathbbm{1}}_{\boldsymbol{d}_{i}} \in \mathcal{P}_{i}^{\boldsymbol d,T},\boldsymbol s^1 \in \boldsymbol S^1, 
\end{align*}
completing the verification.


\end{document}